\documentclass[a4paper,12pt,fleqn]{article}
\usepackage{head,fullpage}  

\usepackage[colorlinks,linkcolor=blue,citecolor=blue]{hyperref}

\usepackage{graphicx}
\usepackage{amsmath,amsfonts,amsthm,amssymb}
\usepackage[dvipsnames]{xcolor}
\usepackage[normalem]{ulem}
\usepackage{breqn}
\usepackage{titlesec}
\usepackage{enumerate}
\usepackage{tikz}
\usepackage{subcaption}
\usepackage[labelfont=bf]{caption}
\usepackage{cite}
\usepackage[titletoc]{appendix}
\usepackage{empheq}

\titleformat*{\section}{\large\bfseries\sffamily}
\titleformat*{\subsection}{\bfseries\sffamily}
\titleformat*{\subsubsection}{\itshape}

\numberwithin{equation}{section}

\newcommand{\del}{\partial}
\newcommand{\C}{\mathcal{C}}

\newcommand{\avg}[1]{\left\langle#1\right\rangle}

\newcommand{\emil}[1]{#1}

\begin{document}	
	
{\noindent\large\bf\sffamily A comparison of dynamical fluctuations of biased diffusion and run-and-tumble dynamics in one dimension}
	
\begin{minipage}{0.15\textwidth}
\end{minipage}\hfill
\begin{minipage}{0.85\textwidth}
	\vspace{8mm}
	\textbf{\sffamily Emil Mallmin, Richard A Blythe, and Martin R Evans}
	\vspace*{2mm}\newline
	\textit{SUPA, School of Physics and Astronomy, University of Edinburgh,\\ Peter Guthrie Tait Road, Edinburgh EH9 3FD, United Kingdom}
	\vspace*{2mm}\newline
	{\small e-mail: \texttt{emil.mallmin@ed.ac.uk} }
		
	\vspace*{4mm}
	August 20, 2019\ --- revised version
		
	\vspace{12mm}
	\textbf{\sffamily Abstract}\newline
We compare the fluctuations in the velocity and in the fraction of time spent at a given position for minimal models of a passive and an active particle: an asymmetric random walker and a run-and-tumble particle in continuous time and on a 1D lattice. 
We compute rate functions  and  
effective dynamics conditioned on large deviations for these observables.  While generally different, for a unique and non-trivial choice of rates (up to a rescaling of time) the velocity rate functions for the two models become identical, whereas the effective processes generating the fluctuations remain distinct. This equivalence coincides with a remarkable parity of the spectra of the processes' generators. For the occupation-time problem, we show that both the passive and active particles undergo a prototypical dynamical phase transition when the average velocity is non-vanishing in the long-time limit.
\end{minipage}
	
\vfill
	
{Published in \textit{J.\ Phys.\ A:\ Math.\ Theor.:} \url{https://doi.org/10.1088/1751-8121/ab4349}

\thispagestyle{empty}	
\newpage
\setcounter{page}{1}                            
\rhead{\footnotesize\sffamily Dynamical fluctuations of biased diffusion and run-and-tumble dynamics in 1D}
\cfoot{}
\rfoot{\thepage}	
		
\section{Introduction}

A core theme of statistical physics is the characterization of fluctuations in physical observables both in the vicinity of and far away from their typical values. Large deviation theory \cite{Touchette2009} provides quantitative answers to such questions. In recent years, this theory has been applied to nonequilibrium processes and fluctuations in density profiles and particle currents\cite{Derrida2007}, the number of state transitions (i.e.\ dynamical activity) \cite{Jack2010,Jack2013}, entropy production \cite{Speck2012}, and more generally dynamical observables $\mathcal{O}_t$ which integrate some quantity over a long time-interval $t$ \cite{Lecomte2007,Touchette2018,Garrahan2009,Jack2010,Jack2015,Chetrite2013,Chetrite2014}. A common situation is that, given an underlying stochastic dynamics, $\mathcal{O}_t$ satisfies the \textit{large deviation principle}
\begin{equation}\label{eq:LDP_rho}
\text{Prob}(\mathcal{O}_t / t = \rho) = e^{- I(\rho) t + o(t)} \quad \mbox{as $t\to\infty$} \;.
\end{equation}
The \emph{rate function} $I(\rho) \geq 0$ quantifies the likelihood of the observable taking a particular value in the long-time limit: larger values of $I(\rho)$ indicate a lower likelihood, while $I(\rho)=0$ determines the one or more possible stationary values of $\rho$. 

In this work, we compare two paradigmatic Markov jump processes---the simple asymmetric random walker (ARW) and the asymmetric run-and-tumble particle (RTP)---for the fluctuations in two elementary dynamical observables---the total particle displacement, and the time spent at a given position. The corresponding time-intensive quantities ($\mathcal{O}_t/t$ for large $t$) are the asymptotic velocity and occupation time-fraction (occupation, for short); the setting is continuous time and an infinite one-dimensional (1D) lattice. A key difference between the models is that the ARW is a \textit{passive} process whereas the RTP is {\em active} in the sense of  possessing an internal state deciding its direction of motion. As a minimal model for persistent random motion, RTPs have been studied extensively to reveal a wealth of nonequilibrium phenomena for both single and interacting particles \cite{Tailleur2008,Thompson2011,Slowman2016a,Slowman2017,Malakar2018,Dhar2018,Mallmin2019a}. We make the comparison of the RTP and ARW more direct (and generalize parts of our analysis) by viewing them as examples drawn from a broader class of multi-state random walks \cite{Weiss1983,Weiss1994} described in \autoref{sec:model}.

In the four problems we consider---two models and two observables---we calculate the respective rate functions, and, in addition, the transition rates of the corresponding \textit{effective process}. This effective process\footnote{This process is variously referred to as the \emph{effective}, \emph{driven} \cite{Chetrite2014}, \emph{auxiliary} \cite{Jack2010}, or \emph{conditioned} process.}  \cite{Jack2010,Jack2015,Chetrite2014,SzavitsNossan2017} \emil{generalizes the conditioned Brownian motion construction of Doob \cite{Doob1957}, and tells us \textit{how} a chosen, possibly rare, fluctuation of the original process is most likely to be realized. So far, the rather recent effective process formalism \cite{Jack2010} has been applied to the asymmetric simple exclusion process (ASEP) \cite{Popkov2010}, zero-range processes \cite{Harris2013,Hirschberg2015}, kinetically constrained models \cite{Jack2013}, and birth-death processes \cite{Torkaman2015,Torkaman2017}.} In the interest of a self-contained presentation, we briefly recapitulate the mathematical theory of rate functions and the effective process in \autoref{sec:theory}.

Our main result with regard to the velocity fluctuations (\autoref{sec:disp}) is the novel finding that the rate functions for the ARW and the RTP \emph{exactly} coincide for a non-trivial and essentially unique choice of the model parameter values. Remarkably, these values do not render the two processes isomorphic: the respective effective processes demonstrate that the same velocity fluctuation is generated by different kinds of trajectories in the two models. Whereas the existence of a well-defined rate function and its qualitative features, such as convexity, or presence of singularities, are subject to universality, its exact form generally depends on model details. Furthermore, a correspondence between the two models' generator spectra emerges at this parameter tuning, which suggests a deeper connection between the models at this parameter tuning.

Our treatment of occupation fluctuations in \autoref{sec:res} complements and extends the analysis of Tsobgni Nyawo and Touchette \cite{Nyawo2016,Nyawo2017,Nyawo2018}. They considered the fraction of time $\rho$ spent by a biased 1D Brownian particle in a given finite interval, which constitutes a simple prototype for a dynamical phase transition, in the sense of a non-analytic rate function. As they show, a critical value $\rho_c$ separates a high-occupation regime generated by trajectories that stay localized around the target interval for their whole duration, and a low-occupation regime generated by trajectories which terminally escape the interval after some time-fraction. Our lattice formulation of this problem (the ARW) reproduces the phenomenology of Ref.\ \cite{Nyawo2018}, with the additional benefit of being solvable in closed form for the rate function and effective rates. For the class of walkers we set out, the occupation rate function can be obtained via the leading singularity of a dynamical partition function in Laplace space, for which we derive a formula which can be solved with numerical exactness. Thus, we are able to demonstrate that the phase transition exist also in the RTP if and only if the asymptotic velocity is non-zero, just as in the ARW case. This lends support to a conjecture of Ref.\ \cite{Nyawo2018}, that one should expect the presence of this transition for random walks possessing two main qualities. The original process must be transient rather than recurrent, i.e.\ with probability one the particle revisits its starting point only finitely often; or at least it allows transience to emerge upon conditioning on the large deviations if originally it was recurrent. In 1D, recurrence amounts to zero asymptotic velocity. Furthermore, the \emph{tilted} process generator, which deforms the original generator through a continuous parameter $s$ (see \autoref{sec:theory}), must allow eigenvalue crossings in $s$. This cannot be the case for an original equilibrium process, whose tilted generator is necessarily symmetrizable and hence avoids crossings \cite{Jack2010}.

A number of previous studies contain calculations related to what we present here. The velocity fluctuations for the ARW were first derived by Lebowitz and Spohn \cite{Lebowitz1999}, and have been extended to inhomogeneous space in, e.g., Refs.\ \cite{Nyawo2016b,Masharian2018,Proesmans2019a}. \emil{For two or more ARWs interacting by hardcore exclusion, i.e.\ the ASEP, the effective process yielding non-local interactions was studied by Popkov \textit{et al} \cite{Popkov2010}. Dynamical fluctuation analyses of two-state processes similar to our RTP model appear in Refs.\ \cite{Pietzonka2016,Aurell2017,Murugan2017,Whitelam2018,Proesmans2019,Gradenigo2019}. Proesmans \textit{et al} study persistent walks, on- and off-lattice, and in arbitrary dimension $d$. In the symmetric and driven off-lattice case, they discover a transition to ballistic motion for $d > 5$. Their analysis of the 1D on-lattice RTP differs slightly from ours, which is continuous-time and asymmetric. Moreover, we derive the effective process for the RTP and use it to explain the flatness of the rate function at large time-scale separations of the model dynamics \cite{Pietzonka2016}. Beyond the single-particle problem of Ref.\ \cite{Nyawo2016}, the large deviations of occupation statistics of interacting particles were recently the subject of Ref. \cite{Agranov2019}.}

Finally, it is worth noting that scaling forms different than \eqref{eq:LDP_rho} exist. For instance, joint scaling limits may be considered, such as a long-time limit taken simultaneously with a low-noise \cite{Escamilla2019} or large system-size limit \cite{Maes2008}. Different scaling forms, in which the exponent of time is different from one, may be found in different parameter regimes within the same model \cite{Cagnetta2017,Nickelsen2018,Gradenigo2019}. Gradenigo and Majumdar \cite{Gradenigo2019} studied a non-standard off-lattice RTP in an acceleration field, and derived distinct scaling forms of the long-time velocity distributions in different field-strength regimes.  For dynamical (phase) transitions, where the rate function is non-analytic at some conditioning value, see e.g.\ Refs.\ \cite{Garrahan2009,Speck2012,SzavitsNossan2014,SzavitsNossan2017,Nyawo2018}. Alternatively, a model control parameter, e.g.\ a transition rate tending to zero, may produce singularities \cite{Whitelam2018,Pietzonka2016}. The minimal conditions under which dynamical transitions arise have not yet been pin-pointed, nor is a full taxonomy of possible dynamical fluctuation phenomena available; hence our interest in a comparative study of minimal models.

\color{Black}

\section{Model description}\label{sec:model}


\begin{figure}
	\centering
	\begin{subfigure}{.45\textwidth}
		\centering
		\includegraphics[trim=0 -0.8cm 0 -.8cm,clip,scale=0.6]{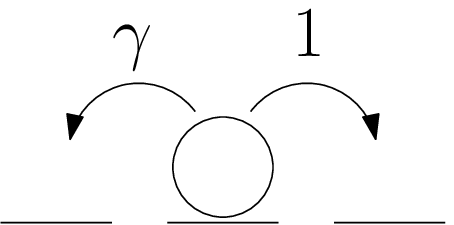}
		\caption{Asymmetric random walker (ARW)}\label{fig:model_arw}
	\end{subfigure}%
	\hspace{1cm}%
	\begin{subfigure}{.45\textwidth}
		\centering
		\includegraphics[scale=0.6]{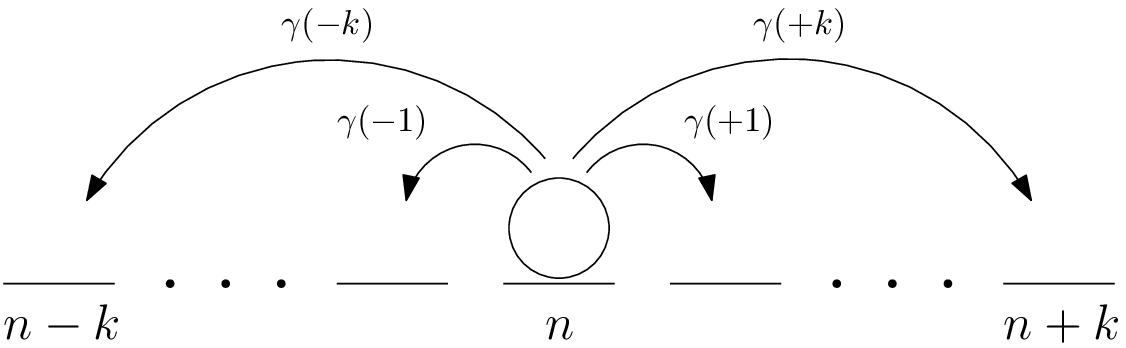}
		\caption{A multi-step walker}
	\end{subfigure}
	\\[0.5cm]
	\begin{subfigure}{.45\textwidth}
		\centering
		\includegraphics[trim=0 -1.2cm 0 -1.2cm,clip,scale=.6]{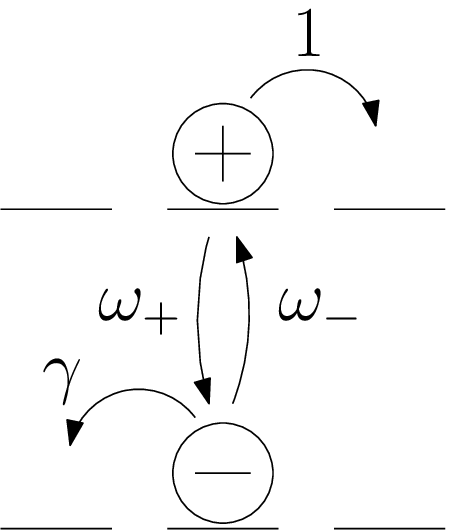}
		\caption{Asymmetric run-and-tumble particle (RTP)}\label{fig:model_rtp}
	\end{subfigure}%
	\hspace{1cm}%
	\begin{subfigure}{.45\textwidth}
		\centering
		\includegraphics[scale=0.6]{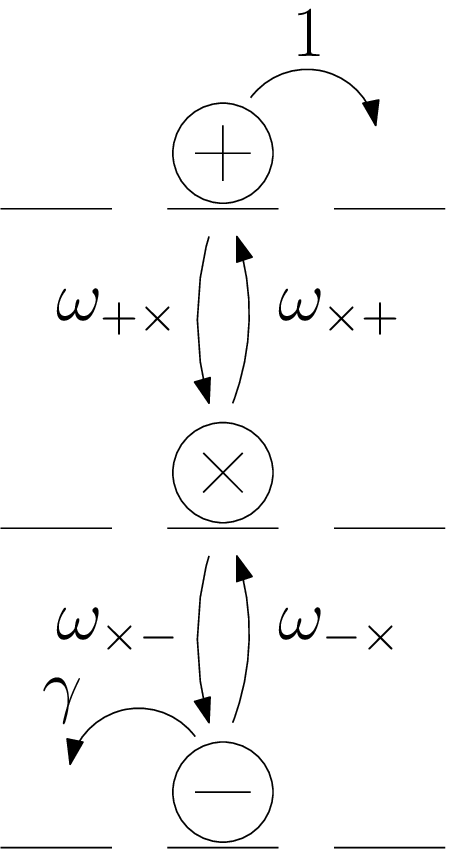}
		\caption{A three-state model: run-and-tumble particle with a refractory state.}
	\end{subfigure}	
	\caption{Examples from the class of models described by the master equation \eqref{eq:ME_class}.}\label{fig:model}
\end{figure}
In this section introduce out a general class of multi-state random walkers in continuous time and on a one-dimensional lattice. The probability distribution on configurations $\C = (i,n)$, comprising the internal state $i$ and the lattice site $n$, evolves according to the master equation
\begin{equation}\label{eq:ME}
\del_t P(\C,t) = \sum_{\C'} M(\C,\C')P(\C',t) \;.
\end{equation}
The Markov matrix, i.e.\ forward generator of the process, is defined in terms of the transition rates $W(\C\to\C')$ and escape rates $\xi(\C) = \sum_{\C'\ne\C}W(\C\to\C')$ as 
\begin{equation}\label{eq:M}
M(\C',\C) = W(\C\to\C') - \xi(\C)\delta_{\C,\C'} \;.
\end{equation}
More specifically, when in internal state $i$, the particle makes a step of $\Delta n$ lattice sites (which may be a positive or negative number) with rate $\gamma_i(\Delta n)$. The particle switches internal state from $i$ to $j$ with a rate $\omega_{ji}$ (with $\omega_{ii}=0$). Importantly, we take all rates to be independent of the spatial coordinate $n$ (i.e., the process is translationally invariant). Then
\begin{equation}\label{eq:ME_class}
	\del_t P_i(n) = \sum_{k} \gamma_i(k) [  P_i(n-k) - P_i(n)] + \sum_{j}\left[ \omega_{ij}P_j(n) - \omega_{ji}P_i(n) \right] \;.
\end{equation}
Examples of models contained in this class are illustrated in \autoref{fig:model}.

To solve the master equation (\ref{eq:ME_class}), we introduce the generating function
\begin{equation}\label{eq:g_i}
	g_i(z,t) = \sum_{n=-\infty}^\infty z^n P_i(n,t) \;.
\end{equation}
Taking $z=e^{iq}$ retrieves a Fourier series, which  has the inversion 
\begin{equation}\label{eq:inv}
P_i(n) = \frac{1}{2\pi i}\oint_{\del D} \frac{dz}{z^{n+1}} g_i(z),
\end{equation}
where $D$ is the complex unit disc. Using vector notation, $\boldsymbol{g} = (g_1,\ldots,g_N)^\top$, where $N$ is the number of internal states. The transform of Eq.~\eqref{eq:ME_class} can be written succinctly as
\begin{equation}\label{eq:Fourier_ME}
\del_t \boldsymbol{g}(z,t) = W(z) {\boldsymbol{g}}(z,t) \;.
\end{equation}
Here, the $N\times N$ matrix $W(z)$ has elements
\begin{subequations}\label{eq:Wz_org}
\begin{align}
W_{ii}(z) &= \sum_k \gamma_i(k)(z^k - 1) - \sum_{j} \omega_{ji}\;,\\
W_{ij}(z) &= \omega_{ij}\quad (i\neq j)\;.
\end{align}
\end{subequations}
The simple form of these equations arises from the translational invariance of the process.

In this work we focus on two instances from this general class of models. The first is the \emph{asymmetric random walker} (ARW). It has a single internal state, and hopping rates according to \autoref{fig:model_arw}. Without loss of generality we choose units of time such that the right hopping rate is unity and the left hopping rate is $\gamma \leq 1$.
The master equation reads
\begin{gather}
\del_t P(n) = P(n-1) + \gamma P(n+1) - (1+\gamma)P(n)\;,\\
W(z) = z + \gamma z^{-1} - (1+\gamma)\;.\label{eq:WARW}
\end{gather}

The second model instance is the \emph{run-and-tumble particle} (RTP). It has two internal states, denoted $+$ and $-$, which indicate the allowed hopping direction when in that state. This model thus introduces persistence of motion, which can be thought of as inertia or self-propulsion. As shown in \autoref{fig:model_rtp}, the particle hops to the right with rate $1$ when in the $+$ state, and to the left with rate $\gamma \leq 1$ (same parameter as the ARW) when in the $-$ state. Switching from the $+$ state to the $-$ state occurs at rate $\omega_+$, and from $-$ to $+$ at rate $\omega_{-}$. In the bacterial system that serves as inspiration for this model \cite{Berg2004}, this velocity-switching is referred to as \emph{tumbling} and we shall use this term in the present work. Given these definitions, the master equation for the RTP is
\begin{subequations}
	\begin{align}
		\del_t P_+(n) &= P_+(n-1) + \omega_- P_-(n) - (1+\omega_+) P_+(n)\;, \\
		\del_t P_-(n) &= \gamma P_-(n+1) + \omega_+ P_+(n) - (\gamma+\omega_-) P_-(n)\;,
	\end{align}
\end{subequations}
\vspace{-.8cm}
\begin{equation}\label{eq:W_RTP}
W(z) = \begin{pmatrix}
z - 1 - \omega_+ & \omega_- \\
\omega_+ & \gamma (z^{-1} - 1) - \omega_-
\end{pmatrix} \;.
\end{equation}
In the degenerate limit $\omega_-=\omega_+ \to \infty$, the RTP becomes identical to an ARW with left and right rates $\gamma/2$ and $1/2$. If instead  $\omega_+\to 0$ and $\omega_- \to \infty$ the RTP becomes the above ARW with $\gamma = 0$.

\section{Dynamical large deviation theory}\label{sec:theory}
We here set out the main ideas behind dynamical large deviation theory in the context of time-homogeneous Markov jump processes. The purpose of the theory is to characterize the likelihood of a given fluctuation in a time-additive observable, and to inform us as to which trajectories typically realize that fluctuation. We refer the reader to Refs.\ \cite{Garrahan2009,Jack2010,Chetrite2014,Touchette2018} for a more complete account.

Over a time $t$, a Markov jump process generates a sequence of configurations wherein a transition from configuration $\C_{i-1}$ to $\C_i$ takes place at time $t_i$, with $0<t_1<\dots<t_n < t$. Together, the sequence of configurations and jump times constitute a trajectory $\mathcal{T}_t$. We consider the class of dynamical observables $\mathcal{O}[\mathcal{T}_t]$ which are time-additive:  $\mathcal{O} = \mathcal{A}+\mathcal{B}$, with
\begin{subequations}
\begin{align}\label{eq:Atype}
\mathcal{A}[\mathcal{T}_t] &= \sum_{i=0}^{n-1} \alpha(\C_i\to\C_{i+1})\;,\\
\mathcal{B}[\mathcal{T}_t] &= (t-t_n)b(\C_n) + \sum_{i=0}^{n-1} (t_{i+1}-t_i)b(\C_i)   = \int_0^t dt'\,b(\C(t')) \;.\label{eq:Btype}
\end{align}
\end{subequations}
The $\mathcal{A}$-part implements a weighted transition counting via the function $\alpha(\C\to\C')$, and the $\mathcal{B}$-part measures the total amount of some quantity continually produced with a state-dependent rate $b(\C)$.

\subsection{The rate function}\label{sec:rf}

The likelihood of a fluctuation is quantified by the rate function appearing in Eq.~\eqref{eq:LDP_rho}, more precisely defined by
\begin{equation}
I(\rho) = -\lim_{t\to\infty}\frac{1}{t} \ln P(\rho,t) \;,
\end{equation}
where 
\begin{equation}\label{eq:basic_LDP}
P(\rho,t) =  \int \mathcal{D}[\mathcal{T}_t]\, \text{Prob}[\mathcal{T}_t] \delta(\mathcal{O}[\mathcal{T}_t] - \rho t) \;.
\end{equation}
Eq.\ \eqref{eq:basic_LDP} invokes the notion of a dynamical ensemble: a probability measure over the set of trajectories generated by \eqref{eq:ME}.  In direct analogy with equilibrium statistical mechanics, one can construct microcanonical and canonical ensembles---with trajectories replacing equilibrium microstates, and the dynamical observable replacing energy. That is, in the microcanonical ensemble, the value of the dynamical observable $\rho$ is considered fixed, whereas in the canonical ensemble it may fluctuate through exchange with some `reservoir' at fixed `inverse temperature' $s$. In this canonical ensemble, we can define the (dynamical) partition function as\begin{equation}\label{eq:Z}
Z(t,s) = \int \mathcal{D}[\mathcal{T}_t]\, \text{Prob}[\mathcal{T}_t]e^{-s\mathcal{O}[\mathcal{T}_t]} = \avg{e^{-s\mathcal{O}}}_t \;.
\end{equation}
Similar to a free energy density, the \textit{scaled cumulant generating function} (SCGF) is defined from \eqref{eq:Z} as
\begin{equation}\label{eq:lambda_lim_Z}
\Lambda(s) = \lim_{t\to\infty} \frac{1}{t} \ln Z(t;s) \;.
\end{equation}
\emil{A central result of large deviation theory, the G\"artner-Ellis theorem (GE) \cite{Touchette2009}, holds that if $\Lambda(s)$ exists and is differentiable in $s$, then the large deviation principle \eqref{eq:basic_LDP} is valid, and the rate function is the Legendre-Fenchel (LF) transform of the SCGF:
\begin{equation}\label{eq:I}
I(\rho) = \max_{s} \{ - \Lambda(s) - s\rho  \} \;.
\end{equation}
In our two velocity problems, the GE conditions are satisfied, whereas in the occupation problems, we find the SCGF to have a non-differentiable point. In that case, we invoke Varadhan's theorem \cite{Touchette2009}, which states that if the large deviation principle is assumed to hold, then  
\begin{equation}\label{eq:varadhan}
\Lambda(s) = \max_\rho \{ - I(\rho) - s\rho \}\;.
\end{equation}
The LF transform appearing in both \eqref{eq:I}  and \eqref{eq:varadhan} always produces a convex function, and is its own inverse transformation if applied to convex functions. Hence, if we can motivate the existence and convexity of the rate function, which we will do by direct numerical simulation, then \eqref{eq:I} follows by inverting \eqref{eq:varadhan}. Otherwise, if the true rate function is non-convex, such as when there are several possible stationary limits (see e.g.\ \cite{Whitelam2018}), \eqref{eq:I} would only yield the convex envelope of the true rate function.} 

To find $\Lambda(s)$, we return to the master equation \eqref{eq:ME} to define a new \textit{tilted} generator
\begin{equation}\label{eq:M(s)}
M(\C',\C;s) = W(\C\to\C')e^{- s \alpha(\C\to\C')} - (\xi(\C) + s b(\C) )\delta_{\C,\C'} \;.
\end{equation}
This generator defines the time evolution of the `probability' $P(\C,t;s)$ which is not conserved, and is merely a computational aid. Its usefulness stems from the fact that \cite{Garrahan2009}
\begin{equation}\label{eq:Zdef2}
Z(t,s) = \sum_{\C} P(\C,t;s)\;.
\end{equation}
Therefore, if the spectrum of $M(s)$ \eqref{eq:M(s)} is gapped, one can conclude via \eqref{eq:lambda_lim_Z} that $\Lambda(s)$ is the dominant eigenvalue (i.e.\ with largest real part) of $M(s)$, since $P(\C,t;s)\sim e^{\Lambda(s)t}$ for large $t$. By the Perron-Frobenius theorem, this eigenvalue is real and simple (at least on finite lattices). 

In the net displacement problem we obtain $\Lambda(s)$ directly from $M(s)$, whereas in the occupation fluctuation problem we take the route of determining the Laplace-transform $\tilde{Z}(u,s)$ of $Z(t,s)$. It then suffices to determine the location of its singularity $u^*(s)$ that lies furthest to the right in the complex-$u$ plane. Then,  for large $t$, $Z(t,s) \sim {\rm e}^{u^*(s) t}$, and hence from \eqref{eq:lambda_lim_Z} we have that $\Lambda(s)=u^*(s)$. Thus, for our multi-state random walkers where $\C=(i,n)$, it proves useful to define
\begin{equation}
Z_i(t,s) = \sum_{n=-\infty}^{\infty} P_i(n,t;s).
\end{equation} 
Writing in vector notation $\boldsymbol{Z} = (Z_1,Z_2,\ldots,Z_N)^\top$ and  recalling that $N$ is the number of internal states, we find $\boldsymbol{Z}$ to be related to the generating function $\boldsymbol{g}(z,t;s)$ of the tilted process by
\begin{equation}
\boldsymbol{Z}(t,s) = \boldsymbol{g}(1,t;s) \;,
\end{equation}
and to $Z$ via
\begin{equation}\label{eq:Zvec2}
Z(t,s) = \boldsymbol{1}^\top \boldsymbol{Z}(t,s) \;,
\end{equation}
where $\boldsymbol{1}$ is the appropriate column vector of ones. 

\subsection{The effective process}\label{sec:eff}

To the question of which trajectories realize a certain fluctuation, the exact answer is given by the microcanonical distribution, $\text{Prob}[\mathcal{T}_t\,|\,\mathcal{O}[\mathcal{T}_t] = \rho t]$. This distribution is difficult to find directly. It is however possible to construct a time-homogeneous \textit{effective} Markov process \cite{Jack2010,Chetrite2014} which represents the same long-time stochastic dynamics (in a sense discussed below). For any value of the parameter $s$, the effective process has rates
\begin{equation}\label{eq:Wd}
W^{\text{eff}}(\C\to\C';s) = e^{-[V(\C';s) - V(\C;s)] - s\alpha(\C\to\C')}  W(\C\to\C')\;.
\end{equation}
The potential $V(\C;s) = - \log \ell(\C;s)$ is defined from the \textit{left}\footnote{\emil{Note that other references may use the \textit{right} eigenvector of the (backward) generator $L$, which is the adjoint/transpose of $L^\dagger = M$.}} eigenvector $\ell(\C;s)$ corresponding to the dominant eigenvalue $\Lambda(s)$ of the matrix $M(s)$ defined by \eqref{eq:M(s)}. The transition counting $\alpha(\C\to\C')$ enters the effective rate as a nonequilibrium driving force since it cannot in general be expressed as a potential difference. We usually require the normalization 
\begin{equation}\label{eq:norm}
\sum_\C \ell(\C)r(\C) = 1 \;,
\end{equation}
with $r$ the matching right eigenvector, as $\ell(\C) r(\C)$ gives the steady-state probability distribution of the effective process (if it exists) \cite{Garrahan2009,Touchette2018}.

By construction, if $I(\rho)$ is \textit{strictly} convex and differentiable at $\rho$, the choice $s=-I'(\rho)$ (i.e.\ the maximizer of \eqref{eq:I}) makes the fluctuation $\rho$ of the original process typical under the effective process. \emil{Furthermore, the dynamical ensemble of this effective process becomes in an asymptotic sense equivalent to the microcanonical ensemble constrained on achieveing $\rho$ (see \cite{Chetrite2014} for a mathematically precise statement). If the SCGF has singularities, the rate function either has a linear or non-convex part; in any case the asymptotic equivalence of effective and microcanonical dynamical ensembles breaks down in this regime. It may be informative to study how the effective process as a function of $s$ changes qualitatively at the point where the ensemble equivalence fails, and to resort to numerical simulation to determine the structure of trajectories in the non-convex regimes \cite{Nyawo2018}.}

The effective process generally conserves some aspects of the original transition rates \cite{Chetrite2014}. \emil{In our problems, we will check the (non-)preservation of the time-symmetric component of the transition rates}---activity parameters, in the terminology of \cite{Maes2018}---as they relate to the total number of state transitions. Define an operator $\theta$ which acts as kinematic reversal on RTP configurations, $\theta(\pm, n) = (\mp, n)$, and as the identity for ARW configurations. The activity parameters are then defined
\begin{equation}
a(\C,\C') = [W(\C\to\C') W(\theta \C' \to \theta\C)]^{1/2}\;. \label{eq:act}
\end{equation}

\section{Velocity fluctuations}\label{sec:disp}

We now obtain the rate functions and effective processes for the ARW and RTP for the case where  the dynamical observable $\mathcal{O}_t$ is the net displacement of the particle up to time $t$.  Recalling that $\C=(i,n)$ denotes a configuration where the particle is in internal state $i$ at position $n$, the displacement between two configurations is an observable of the $\mathcal{A}$-type \eqref{eq:Atype} with
\begin{equation}
\alpha\left((i,n)\to (i,n+k)\right) = k
\end{equation}
for each pair of configurations with a nonzero transition rate, $\gamma_i(k) > 0$.

Within the generating function formulation, the tilted process satisfies an equation of the form \eqref{eq:Fourier_ME} with a suitably tilted $W$ matrix. The elements of this matrix are
\begin{equation}\label{eq:W(z,s)}
W_{ij}(z;s) = \left\{ \begin{array}{ll}
{\displaystyle \sum_k \gamma_i(k)(z^ke^{-sk} - 1) - \sum_{k\neq i} \omega_{ki}} & \mbox{if $i=j$} \\[1ex]
\omega_{ij} & \mbox{otherwise} 
\end{array} \right. \;.
\end{equation}
Our aim is to find the dominant eigenvalue of the $NL\times NL$ tilted Markov matrix $M$, \eqref{eq:M(s)}. When the lattice size $L$ is finite and periodic, the eigenvalues of this matrix are obtained by collecting together the eigenvalues from the $L$ different $N\times N$ matrices $W(z;s)$ obtained by putting $z = 1,e^{2\pi i /L},\ldots,e^{2(L-1)\pi i/L}$. As the size, $L$, of the lattice does not affect the velocity fluctuations of a single particle, the largest eigenvalue must come from the case $z=1$ (the only allowed value of $z$ for $L=1$). The problem thus reduces to that of finding the largest eigenvalue of the $N\times N$ matrix $W(1;s)$ given by \eqref{eq:W(z,s)}, or equivalently, $W(e^{-s})$ with $W(z)$ given by \eqref{eq:Wz_org}. Since $N$ is finite, the Perron-Frobenius theorem dictates that the dominant eigenvalue $\Lambda(s)$ is simple, and hence does not cross with other eigenvalues as $s$ is varied. Since the matrix elements are differentiable in $s$, so is $\Lambda(s)$, and the G\"artner-Ellis theorem is applicable. 

The asymptotic average velocity $\bar{v}$, for which we will find $I(\bar{v})=0$, can be computed from
\begin{equation}\label{eq:avgv}
\bar{v} = \sum_i P^*_i \bar{v}_i\;,
\end{equation}
where $P^*_i$ is the steady-state distribution over the internal states $i$, and $\bar{v}_i$ is the average velocity of the particle conditioned on being in state $i$,
\begin{equation}\label{eq:avgvi}
\bar{v}_i = \sum_k \gamma_i(k) k \;.
\end{equation}
In particular, for the ARW 
\begin{equation}\label{eq:v_ARW}
\bar{v}^{\text{ARW}} = 1  - \gamma. 
\end{equation}
For the RTP, the steady-state condition of the internal dynamics is 
\begin{equation}
P_+^* \omega_+ = P_-^* \omega_- \: ,
\end{equation}
which yields
\begin{equation}
P_+^* = \frac{\omega_-}{\omega_++\omega_-}\;,\quad
P_-^* = \frac{\omega_+}{\omega_++\omega_-}\; .
\end{equation}
The average velocity is therefore
\begin{equation}\label{eq:v_RTP}
\bar v ^{\text{RTP}} = P_+^* -\gamma P_-^* =  \frac{\omega_--\gamma \omega_+}{\omega_++\omega_-}\;.
\end{equation}

\subsection{Asymmetric random walker---rate function}

The case of the ARW is particularly simple, as the number of internal states $N=1$, and the matrix \eqref{eq:W(z,s)} is a scalar equal to its only (and hence dominant) eigenvalue.  To express this eigenvalue, it is convenient to introduce the functions
\begin{equation}\label{eq:coshg}
\cosh_\gamma(x) := \frac{e^{x} + \gamma e^{-x}}{2}, \quad \sinh_\gamma(x) := \frac{e^{x} - \gamma e^{-x}}{2} \;.
\end{equation}
Then, from \eqref{eq:W(z,s)} with $z=1$ we have
\begin{equation}
\Lambda(s) = W_{11}(1;s) = 2[\cosh_\gamma(-s) - \cosh_\gamma(0)] \;.
\end{equation}
Putting $\rho=v$, the desired asymptotic velocity, into \eqref{eq:I}, we find the maximizer $s^*$ to be
\begin{equation}\label{eq:ARW_vel_s*}
\sinh_\gamma(-s^*) = v/2\quad \iff\quad 
s^*  = \ln \left[ \frac{1}{\gamma}\left( \sqrt{\gamma + (v/2)^2} - v/2  \right) \right].
\end{equation}
Using $\cosh_\gamma(x) = (\gamma + \sinh_\gamma^2(x))^{1/2}$, we find  the rate function
\begin{equation}\label{eq:I_vel_ARW}
I_\gamma(v) =  {1+\gamma} - 2\sqrt{\gamma+(v/2)^2} - v\ln \left[ \frac{1}{\gamma}\left( \sqrt{\gamma + (v/2)^2} - v/2\right) \right] \;.
\end{equation}
This result, previously obtained in e.g.~\cite{Lebowitz1999,Speck2012,Masharian2018}, is plotted in \autoref{fig:vel_rf}.

\subsection{Asymmetric random walker---effective process}\label{sec:vel_arw_dp}

\begin{figure}[t]
	\centering
	\begin{subfigure}{0.45\textwidth}
		\begin{tikzpicture}
		\node[anchor=south west,inner sep=0] (image) at (0,0){\includegraphics[scale=1.2]{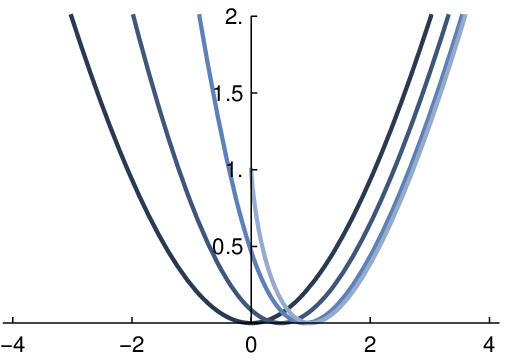}};
		\begin{scope}[x={(image.south east)},y={(image.north west)}]
		\node at (0.5,1.1) {$I_\gamma(v)$};
		\node at (1.05,0.09) {$v$};
		\node at (0.05,.9) {$\gamma = 1$} ;
		\node at (0.24,0.81) {$0.5$} ;
		\node at (0.39,0.7) {$0.1$} ;
		\node at (0.55,0.48) {$0$} ;
		\end{scope}
		\end{tikzpicture}
		\caption{Rate function $I_\gamma(v)$ plotted for $\gamma = 1,0.5,0.1$ and 0. The same rate function is found for the RTP if $\omega_+=\gamma$ and $\omega_-=1$.}\label{fig:vel_rf}
	\end{subfigure}%
	\hspace{1cm}%
	\begin{subfigure}{0.45\textwidth}
		\begin{tikzpicture}
		\node[anchor=south west,inner sep=0] (image) at (0,0){
			\includegraphics[scale=1.2]{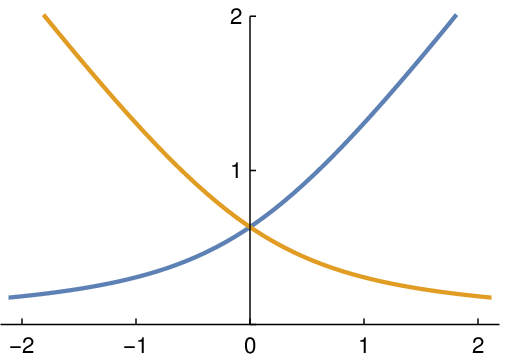}};
		\begin{scope}[x={(image.south east)},y={(image.north west)}]
		\node at (0.66,0.37) {$\sqrt\gamma$};
		\draw[->] (0.6,0.37)--(0.54,0.37);
		\node at (0.28,0.86) {$\gamma^{\text{eff}}(-1)$};
		\node at (0.72,0.86) {$\gamma^{\text{eff}}(+1)$};
		\node at (1.05,0.08) {$v$};
		\end{scope}
		\end{tikzpicture}
		\caption{Hopping rates for the effective process conditioned on obtaining limiting velocity $v$. }\label{fig:vel_rates_ARW}
	\end{subfigure}
	\caption{Velocity fluctuations of the ARW}
\end{figure}

The rates of the effective process, \eqref{eq:Wd}, involve the left eigenvector of the tilted $M$ matrix corresponding to the dominant eigenvalue. Since in this case the problem reduces to a $1\times 1$ eigenvalue problem, it follows that the left eigenvector is uniform over all configurations, and one quickly finds that
\begin{subequations}
\begin{align}
	W^{\text{eff}}(n\to n+1)  &= e^{-s}\\
	W^{\text{eff}}(n+1\to n) &= \gamma e^{s}.
\end{align}
\end{subequations}
Substituting \eqref{eq:ARW_vel_s*} for $s$ produces the following rates, plotted in \autoref{fig:vel_rates_ARW},
\begin{equation}
	W^{\text{eff}}(n\to n\pm 1) = \gamma^{\text{eff}}(\pm 1) =  \sqrt{\gamma + (v/2)^2} \pm v/2\;.
\end{equation}

Naturally, the effective process has preserved the spatial homogeneity of the original process. \emil{There, the activity parameters were constant across all configurations. When this is the case, the activity parameters are proportional to the traffic, i.e.\ total number of (undirected) jumps on the lattice \cite{Maes2019}.} As the effective process has preserved the values of the activity parameters, an atypical velocity $v$ of the original process apparently tends to arise through a trajectory where the total \emph{number} of jumps (irrespective of direction) is the same as for the case of the typical velocity $\bar{v}$, but where the relative probability of jumps in each direction is modified to obtain the desired $v$. Intuitively, a trajectory achieving $v$ with an atypically large (or small) number of jumps can be made more probable if an equal number of left and right jumps are removed (or added) to restore typical levels of traffic. This explains the preservation of activity parameters in this case.

\subsection{Asymmetric run-and-tumble particle---rate function}

\begin{figure}[t]
	\centering
	\begin{subfigure}{0.45\textwidth}
		\begin{tikzpicture}
		\node[anchor=south west,inner sep=0] (image) at (0,0){\includegraphics[scale=1.2]{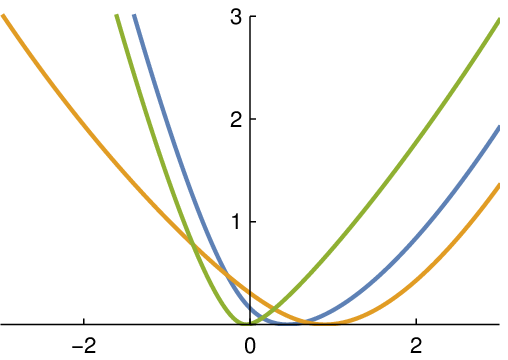}};
		\begin{scope}[x={(image.south east)},y={(image.north west)}]
		\node at (0.5,1.1) {$I(v)$};
		\node at (1.05,0.09) {$v$};
		\node[align=left] at (0.88,1.05) {$(\gamma, \omega_{\pm}) =$};
		\node[align=left] at (1.11,0.95) {$(0.1,1)$};
		\node[align=left] at (1.165,.68) {$(1,1\mp0.9)$};
		\node[align=left] at (1.20,0.48) {$(0.1,1\pm 0.9)$};
		\end{scope}
		\end{tikzpicture}
		\caption{Rate function for combinations of symmetric/asymmetric hopping and tumbling.}\label{fig:vel_rf_rtp}
	\end{subfigure}%
	\hspace{1cm}%
	\begin{subfigure}{0.45\textwidth}
		\begin{tikzpicture}
		\node[anchor=south west,inner sep=0] (image) at (0,0){
			\includegraphics[scale=1.2,trim=0 0 0 0,clip]{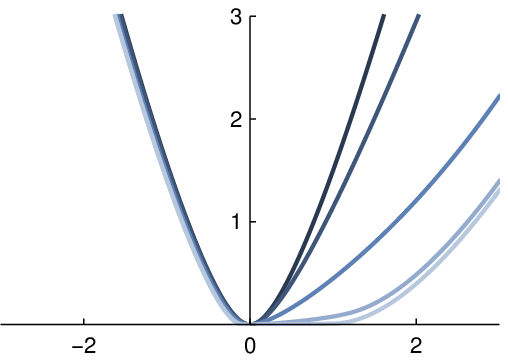}};
		\begin{scope}[x={(image.south east)},y={(image.north west)}]
		\node at (0.5,1.08) {$I(v)$};
		\node at (1.05,0.09) {$v$};
		\end{scope}
		\end{tikzpicture}
		\caption{Rates \eqref{eq:0vrates} giving zero limiting velocity; $\gamma=0.1$, and $\delta = 10^n$, $n=2,1,0,-1,-2$ from dark to light.}\label{fig:vel_rf_delta}
	\end{subfigure}
	\caption{Velocity rate function for RTP}\label{fig:vel_rf_RTP}
\end{figure}

We now perform the equivalent calculation for the RTP. The key difference here is that the matrix $W(z;s)$ defined by \eqref{eq:W_RTP} is two dimensional. When $z=1$, the larger eigenvalue is
\begin{equation}\label{eq:RTPve}
\Lambda(s) = \cosh_\gamma(-s) - \cosh_\gamma(0) - \omega + \sqrt{ \omega^2 - \delta \omega^2 + (\sinh_\gamma(-s) - \sinh_\gamma(0) - \delta \omega)^2} \;,
\end{equation}
recalling the definitions \eqref{eq:coshg} and introducing
\begin{equation}
\omega = \frac{\omega_+ + \omega_-}{2}, \quad \delta \omega = \frac{\omega_+ - \omega_-}{2} \;. 
\end{equation}
The maximizer $s^* = -I'(v)$ satisfies
\begin{equation}\label{eq:v=x...}
v =  x + \frac{ (x - x_0 - \delta\omega) \sqrt{\gamma + x^2}}{\sqrt{\omega^2 - \delta \omega^2 + (x - x_0 - \delta\omega)^2}} \;,
\end{equation}
where
\begin{equation}
x = \sinh_\gamma(-s^*)\quad\mbox{and}\quad x_0 = \sinh_\gamma(0) \;.
\end{equation}
Solving (\ref{eq:v=x...}) for $x$ (and therewith $s^*$) amounts to factorizing a cubic polynomial, which we can always do, but the result is unwieldy. But supposing we have obtained the solution $x=x(v)$, for example by numerical evaluation of the cubic formula, we then have 
\begin{dmath}\label{eq:s*vel_rtp}
s^*  = \ln \left[ \frac{1}{\gamma}\left( \sqrt{\gamma + x(v)^2} - x(v)  \right) \right],
\end{dmath}
and $I(v)$ is easily written down in terms of $x(v)$. We present the solution in the form of a plot of the rate function in \autoref{fig:vel_rf_RTP} for some illustrative parameter values. Some special cases warrant a closer investigation.

\subsubsection{ARW large deviation equivalence} 

If we take the tumbling rates to be
\begin{equation}\label{eq:special_rates}
\omega_+ = \gamma,\quad \omega_- = 1,
\end{equation}
then $\delta\omega + x_0 =0$ and \eqref{eq:v=x...} simplifies to $x(v) = v/2$. Consequently, the rate function $I(v)$ is  \textit{exactly identical} to that of the ARW, Eq.~\eqref{eq:I_vel_ARW}. This \textit{large deviation equivalence} is non-trivial, as evidenced by the following observations: (i) the original processes remain microscopically distinct for this choice of rates (in contrast to the limit of infinite tumble rates); (ii) the effective processes are also microscopically distinct; (iii) the diffusion-limit descriptions (original or effective) are distinct.
\begin{enumerate}[(i)]
	\item As the ARW is a Markov process, the probability of the next jump being to the right is $1/(1+\gamma)$, independently of the direction of the previous jump. The RTP marginalized over the internal states is \textit{not} a Markov process for any (non-trivial) choice of parameters, including \eqref{eq:special_rates}. Consider the probability of the next jump being to the right given that the previous was, $P^{\rm RTP}(+|+)$. We must take into account that the particle may tumble back and forth between the $+$ and $-$ states multiple times before it eventually hops to the right again. Thus, we must sum the probabilities of: not tumbling before hopping to the right; tumbling twice before hopping to the right; four tumbles, and so on. Thus
	\begin{align}
	\nonumber
	P^{\rm RTP}(+|+) &= \frac{1}{1+\omega_+} \left( 1 + \frac{\omega_+}{1+w_+} \frac{\omega_-}{\gamma+ \omega_-} + 
	\left[ \frac{\omega_+}{1+w_+} \frac{\omega_-}{\gamma+ \omega_-} \right]^2 + \cdots  \right) \\&= \frac{\gamma+\omega_-}{\gamma+\omega_-+ \gamma \omega_+} \;.\label{eq:P++RTP}
	\end{align}
	A similar calculation shows that $\displaystyle P^{\rm RTP}(+|-) = 
	\omega_-/(\gamma+\omega_-+ \gamma \omega_+)$. Clearly $P^{\rm RTP}(+|+)$ and $P^{\rm RTP}(+|-)$ can only coincide when $\gamma = 0$. Thus,  for $\gamma>0$  the direction of a hop of the RTP  is never independent  of that of the previous hop, in contrast to the ARW.
	
	\item We derive the respective effective processes in the next section. However, one knows \textit{a priori} that the effective process does not alter the original dynamics qualitatively (as in adding or removing states or transitions). The effective RTP process is therefore, once again, an RTP process which by (i) will be distinct from the effective process of the ARW due to persistence. 
	
	\item On long space and time-scales the two models behave diffusively (assuming non-ballistic parameter choices for the RTP). We introduce a small lattice spacing $a$. The particle density $\rho(x)$ satisfies the drift-diffusion equation
	\begin{equation}
		\del_t \rho = - u \del_x \rho + D \del_x^2 \rho\;.
	\end{equation}
	with $u = a(1-\gamma)$ and $D = a^2(1+\gamma)$ for the ARW, and $u = a(1-\gamma)$ and $D = a^2 \gamma / (1+\gamma)$ for the RTP \cite{Tailleur2008}. Obviously, the limiting velocities are identical, as otherwise the rate functions would not have the same zero, but the diffusion coefficients are different. That this is true also for the diffusive-limit of the effective process, follows from the fact that, as noted in the next section, the RTP effective processes preserves the symmetry of its original rates, and their relation to the rates of the effective ARW process. 	
\end{enumerate}

As mentioned, identical limiting velocities of the ARW and RTP is a necessary condition for the large deviation equivalence, but it is insufficient. Setting  $\bar{v}^{\text{ARW}} = \bar{v}^{\text{RTP}}$, that is \eqref{eq:v_ARW} =  \eqref{eq:v_RTP}, we find
\begin{equation}\label{eq:cond1}
\omega_+ / \omega_- = \gamma\;.
\end{equation}
To obtain \eqref{eq:special_rates} an additional constraint on the rates must be imposed. Consider the constraint that the two processes have the same maximum escape rates,
\begin{equation}\label{eq:cond2}
1+\gamma = \max\{ 1 + \omega_+, \gamma + \omega_- \}.
\end{equation}
Eq.\ \eqref{eq:cond2} together with \eqref{eq:cond1} then imply \eqref{eq:special_rates}, which incidentally also implies a symmetry for the RTP in that \textit{all} configurations have the same escape rate $\xi = 1 + \gamma$, just as for the ARW. It is easily verified that $-\xi$ is in fact an $L$-degenerate eigenvalue of the RTP Markov matrix. (This `macroscopic eigenvalue crossing' was noted in \cite{Mallmin2019a} for the symmetric RTP.) Curiously, the remaining $L$ eigenvalues map perfectly onto the $L$-sized spectrum of the ARW. As demonstrated in  \autoref{fig:spectrum}, these two circumstances are unique to the rates \eqref{eq:special_rates}. Since not only the tilted dominant eigenvalue (the SCGF) coincide for the two models, but essentially the entire original spectrum, one might expect a relationship between the velocity distributions of the models not only in the long-time limit on the exponential scale. We leave a further analysis of this phenomenon for a future work.

\begin{figure}[t]
	\centering
	\begin{subfigure}{0.33\textwidth}
		\begin{tikzpicture}
		\node[anchor=south west,inner sep=0] (image) at (0,0){\includegraphics[scale=1]{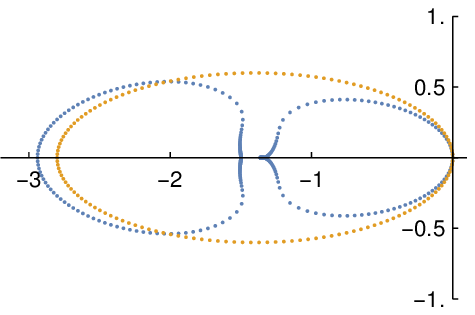}};
		\begin{scope}[x={(image.south east)},y={(image.north west)}]
		\node at (0.07,0.58) {Re};
		\node at (0.85,0.9) {Im};
		\end{scope}
		\end{tikzpicture}
		\caption{$\gamma=0.4$}
	\end{subfigure}
	\begin{subfigure}{0.33\textwidth}
		\begin{tikzpicture}
		\node[anchor=south west,inner sep=0] (image) at (0,0){
			\includegraphics[scale=1]{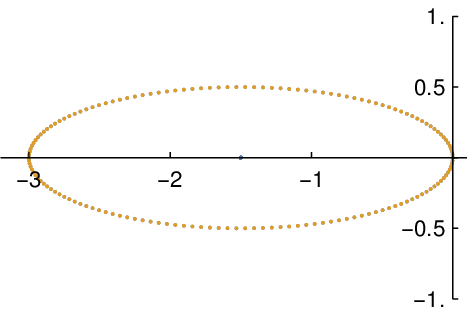}};
		\begin{scope}[x={(image.south east)},y={(image.north west)}]
		\draw[->] (0.51,0.8)--(0.51,0.52);
		\node at (0.51,0.9) {\tiny $L$-fold degeneracy};
		\draw[->] (0.2,0.15)--(0.2,0.3);
		\node at (0.45,.1) {\tiny $L$ overlapping eigenvalue pairs};
		\end{scope}
		\end{tikzpicture}
		\caption{$\gamma=0.5$}
	\end{subfigure}
		\begin{subfigure}{0.33\textwidth}
			\begin{tikzpicture}
			\node[anchor=south west,inner sep=0] (image) at (0,0){
				\includegraphics[scale=1]{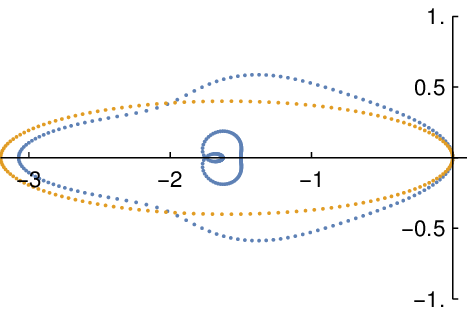}};
			\begin{scope}[x={(image.south east)},y={(image.north west)}]
			\end{scope}
			\end{tikzpicture}
			\caption{$\gamma=0.6$}
		\end{subfigure}
	\caption{\emil{The spectrum in the complex plane for the ARW (orange) and RTP (blue) with $\omega_+ = 0.5$, $\omega_-=1$, $L = 150$, for three values of $\gamma$. In \textbf{(a)} and \textbf{(c)} the spectra are distinct; in \textbf{(b)}, where the RTP rates satisfy \eqref{eq:special_rates}, half the eigenvalues of the RTP coalesce at $-(1+\gamma)$, and the rest reproduce the ARW spectrum.}}\label{fig:spectrum}
\end{figure}

\subsubsection{Zero average velocity}
If we set the tumbling rates to be
\begin{equation}\label{eq:0vrates}
\omega_+ = \delta,\quad \omega_- = \gamma \delta\; ,
\end{equation}
if follows immediately from \eqref{eq:v_RTP} that the average velocity is zero for all values of $\gamma$. As $\delta$ becomes small $I(v)\approx 0$ for $-\gamma < v <1$; see \autoref{fig:vel_rf_delta}. This means that velocities in this entire range are (relatively) likely to be realized. These fluctuations arise from the fact that particles can hop many times in the same direction between velocity reversals. The presence of a widely  fluctuating velocity for random walks with large time-scale separations has been reported on previously \cite{Gradenigo2013,Pietzonka2016,Whitelam2018}. The effective process that we derive in \autoref{sec:RTP_eff} provides another means to investigate this phenomenon.

\subsubsection{ARW limit}
Finally, taking the limit $\delta\omega = 0, \omega\to\infty$ in \eqref{eq:v=x...} we find $x(v) = v $, and the rate function is $I_\gamma(2v)/2$ (compared to \eqref{eq:I_vel_ARW}), which amounts to the ARW of the previous section but with rates rescaled by $1/2$. 

\subsection{Asymmetric run-and-tumble particle---effective process}\label{sec:RTP_eff}

\begin{figure}[t]	\centering
	\begin{subfigure}{0.45\textwidth}
		\begin{tikzpicture}
		\node[anchor=south west,inner sep=0] (image) at (0,0){\includegraphics[scale=1.2]{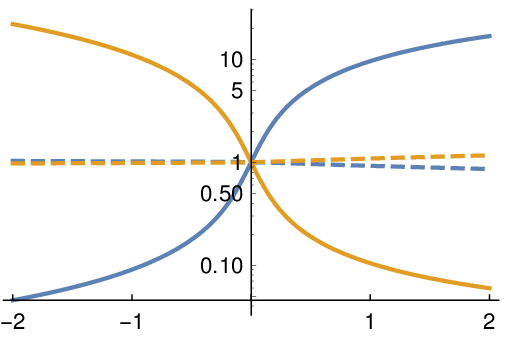}};
		\begin{scope}[x={(image.south east)},y={(image.north west)}]
		\node at (1.04,0.09) {$v$};
		\node at (0.9,0.96) {$\tilde{\gamma}_+$};
		\node at (0.9,0.62) {$\tilde{\omega}_-$};
		\node at (0.9,0.43) {$\tilde{\omega}_+$};
		\node at (0.9,0.22) {$\tilde{\gamma}_-$};
		\end{scope}
		\end{tikzpicture}
		\caption{$\gamma = 0.1$, $\delta = 100$.}\label{fig:delta100}
	\end{subfigure}%
	\hspace{1cm}%
	\begin{subfigure}{0.45\textwidth}
		\begin{tikzpicture}
		\node[anchor=south west,inner sep=0] (image) at (0,0){
			\includegraphics[scale=1.2]{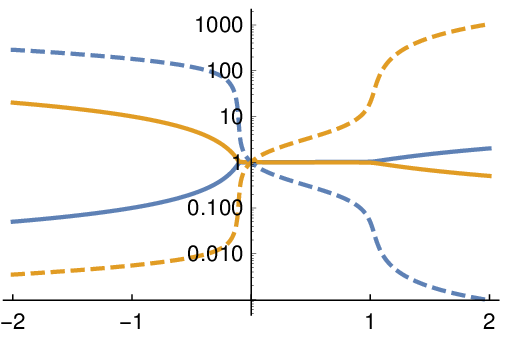}};
		\begin{scope}[x={(image.south east)},y={(image.north west)}]
		\node at (1.04,0.09) {$v$};
		\node at (0.15,0.92) {$\tilde{\omega}_+$};
		\node at (0.15,0.75) {$\tilde{\gamma}_-$};
		\node at (0.15,0.44) {$\tilde{\gamma}_+$};
		\node at (0.15,0.24) {$\tilde{\omega}_-$};
		\end{scope}
		\end{tikzpicture}
		\caption{$\gamma=0.1$, $\delta = 0.01$.}\label{fig:delta0.01}
	\end{subfigure}
	\caption{Rates of the effective process normalized by the original rates ($\tilde{\omega}_+ = \omega_+^{\text{eff}}/\omega_+ $ etc.) plotted on a log-scale versus $v$. Orange (blue) for left (right), unbroken (dashed) line for hopping (tumbling). Parameter values correspond to the rate function \autoref{fig:vel_rf_delta}, i.e.\ tumbling rates according to \eqref{eq:0vrates}. See main text for interpretation.}\label{fig:wonky_rates}
\end{figure}

To calculate the effective process, we require the left eigenvector of the tilted $M$ matrix that corresponds to the eigenvalue \eqref{eq:RTPve}. As for the ARW, the fact that this is obtained from the tilted $W$ matrix with $z=1$ implies that the eigenvector is spatially uniform. However, it has different components, $\ell_+$ and $\ell_-$, for each of the two internal states, the ratio of which can be found from the $2\times 2$ matrix \eqref{eq:W(z,s)} and the eigenvalue \eqref{eq:RTPve}. Then, up to a normalization,
\begin{equation}
\begin{pmatrix} \ell_+ \\ \ell_- \end{pmatrix} = \begin{pmatrix}
x(v) - x_0 - \delta \omega + \sqrt{\omega^2 - \delta \omega^2 +(x(v) - x_0 - \delta \omega)^2} \\
	\omega -\delta\omega
\end{pmatrix}.
\end{equation}
The rates \eqref{eq:Wd} of the effective process realizing the fluctuation $v$ are 
\begin{subequations}
	\begin{gather}
	\gamma_\pm^{\text{eff}} = \sqrt{\gamma + x(v)^2} \pm  x(v) \\
	\omega^{\text{eff}}_{\pm} = \sqrt{\omega^2 - \delta \omega^2 +(x(v) - x_0 - \delta \omega)^2} \mp (x(v) - x_0 - \delta \omega).
	\end{gather}
\end{subequations}
The activity parameter remains unchanged, with the same explanation as in \autoref{sec:vel_arw_dp}.

For the special rates  of the ARW large deviation equivalence, we get the effective rates
\begin{subequations}
\begin{gather}
\gamma_\pm^{\text{eff}} = \sqrt{\gamma + (v/2)^2} \pm  v/2 ,\\
\omega^{\text{eff}}_{\pm} = \sqrt{\gamma + (v/2)^2} \mp v/2.
\end{gather}
\end{subequations}
This shows that the hopping rates are modified in exactly the same way as for the ARW.  In addition to the activity parameter, they retain the property that $\gamma_{\pm}^{\text{eff}} = \omega_{\mp}^{\text{eff}}$ so that the escape rate is the same in every configuration. Nevertheless, the trajectories that generate a velocity $v$ look different from those of the ARW. This is for the same reason as previously noted, that the directions of subsequent hops are correlated in the RTP, whilst they are independent for the ARW.

In \autoref{fig:wonky_rates} we plot the effective rates (normalized by original rates) when the original process satisfies the zero-velocity condition \eqref{eq:0vrates} with $\gamma = 0.1$ (cf.\ \autoref{fig:vel_rf_delta}). For $\delta = 100$, the rate function is close to symmetric around $v=0$, whilst both hopping and tumbling rates are highly asymmetric. From \autoref{fig:delta100} we see that the effective tumbling rates normalized by the original ones remain close to unity whereas the effective hopping rates change considerably with $v$. This explains that a fluctuation $|v|{}> 0$ is caused by trajectories with an atypical proportion of left to right hops, but with typical sequences of tumbles. For $\delta = 0.01$, \autoref{fig:delta0.01} reveals that the near-flat interval $[-0.1,1]$ of the rate function is entirely due to an atypical proportion of left to right tumbles, as the normalized hopping rates remain close to one. On either side of this interval, all four rates are simultaneously modified in the direction of increasing the translational bias. For instance, for $v$ slightly more negative than $-0.1$, the particle drastically increases the tumbling rate from the $+$ to the $-$ state while also increasing the average speed in the $-$ state. The opposite direction of changes is applied to the other two rates, further reducing translation to the right.

\section{Occupation fluctuations}\label{sec:res}

We now turn to the second of the two dynamical observables $\mathcal{O}_t$, the time spent by the walker at a given lattice site, which we take to be site 0. We will refer to the fraction of time $\rho = \mathcal{O}_t/t$ as occupation. We will take site 0 to be the initial site of the particle and further assume that the initial internal state is chosen uniformly. (In fact, the initial condition does not affect the resulting rate function $I(\rho)$, except in the case where a particle can move only in one direction.) $\mathcal{O}_t$ is of the $\mathcal{B}$-type \eqref{eq:Btype}---specifically, $b((i,n)) = \delta_{n,0}$. The tilted process therefore adds a term $- s \delta_{n,0} P_i(0,t)$ to the master equation, and likewise to the generating function equation:
\begin{equation}\label{eq:res_tilted_eq}
\del_t \boldsymbol{g}(z,t;s) = W(z;s) \boldsymbol{g}(z,t;s) - s \delta_{n,0}\boldsymbol{p}(0,t;s),
\end{equation}
with 
\begin{equation}
\boldsymbol{p}(0,t;s) = \frac{1}{2\pi i} \oint_{\del D} \frac{dz'}{z'}\boldsymbol{g}(z',t;s).
\end{equation}
In this instance, the dominant eigenvalue $\Lambda(s)$ of the tilted Markov matrix cannot be straightforwardly related to the eigenvalues of $W(z)$. Instead, we obtain $\Lambda(s)$ from \eqref{eq:lambda_lim_Z}, which tells us that it equals the right-most singularity in the complex plane of the Laplace transformed ($t\to u$) dynamical partition function, $\tilde{Z}(u,s)$. 

To derive a closed formula for $\tilde{Z}$ we begin by taking the Laplace transform $t\to u$ of \eqref{eq:res_tilted_eq} which yields 
\begin{equation}\label{eq:res_Laplace}
u \tilde{ \boldsymbol{g}}(z,u) - \boldsymbol{1} = W(z)\tilde{ \boldsymbol{g}}(z,u) - s \tilde{\boldsymbol{p}}(0,u),
\end{equation}
where the initial condition   
\begin{equation}
\boldsymbol{g}(z,0) = \boldsymbol{Z}(0) = \boldsymbol{1}
\end{equation} 
need not be normalized since `probability' is not conserved. Now, by setting $z=1$ in \eqref{eq:res_Laplace} we find
\begin{equation}\label{eq:res_Z_integral}
u\tilde{\boldsymbol{Z}}(u) - \boldsymbol{1} - \frac{1}{u} \boldsymbol{w} =  -  s\tilde{\boldsymbol{p}}(0,u),
\end{equation}
wherein we have defined
\begin{equation}
(\boldsymbol{w})_i = (W(1)\boldsymbol{1})_i = \sum_{j} (\omega_{ij}-\omega_{ji}).
\end{equation}
We then use \eqref{eq:res_Z_integral} to eliminate $\tilde{\boldsymbol{p}}(0,u)$ in \eqref{eq:res_Laplace}, thereby obtaining after rearrangements
\begin{equation}
(uI - W(z)) \tilde{\boldsymbol{g}}(z,u) = u \tilde{\boldsymbol{Z}}(u) - \frac{1}{u} \boldsymbol{w}.
\end{equation}
Upon inverting the matrix multiplying $\tilde{\boldsymbol{g}}$, and applying  $-(s/2\pi i)\oint dz/z$ on both sides of the equation, we find
\begin{equation}
- \frac{s}{2 \pi i} \oint_{\del D} \frac{dz}{z}\,\tilde{\boldsymbol{g}}(z,u) = - \left[ \frac{s}{2\pi i} \oint_{\del D} \frac{dz}{z}\, (uI - W(z))^{-1}\right] \left(u \tilde{\boldsymbol{Z}}(u) - \frac{1}{u} \boldsymbol{w}\right).
\end{equation}
Recognizing the left hand side as $-s \tilde{\boldsymbol{p}}$ we again eliminate it using \eqref{eq:res_Z_integral}. Rearranging the resulting expression,
\begin{equation}
\left[ I + \frac{s}{2 \pi i} \oint_{\del D} \frac{dz}{z}\,(u I - W(z))^{-1}  \right]\left(u \tilde{\boldsymbol{Z}}(u) - \frac{1}{u} \boldsymbol{w}\right) = \boldsymbol{1}.
\end{equation}
Finally, we invert the expression in square brackets, and apply $\boldsymbol{1}^\top$ from the left. Noting that $\boldsymbol{1}^\top \boldsymbol{w} = 0$ we reach the final result,
\begin{equation}\label{eq:tildeZ}
\tilde{Z}(u,s) = \frac{1}{u}\cdot \boldsymbol{1}^\top \left[ I + s\mathcal{I} \right]^{-1} \boldsymbol{1},\qquad \mathcal{I}=\frac{1}{2 \pi i} \oint_{\del D} \frac{dz}{z}\,(u I - W(z))^{-1}.
\end{equation}
There is in general a pole at $u=0$, so that $\Lambda(s)\geq 0$. As we will see, there is an additional largest positive pole which crosses the zero at some value $s^* \leq 0$.

\subsection{Asymmetric random walker---rate function}

\begin{figure}[t]
	\centering
		\begin{tikzpicture}
		\node[anchor=south west,inner sep=0] (image) at (0,0){\includegraphics[scale=1.2]{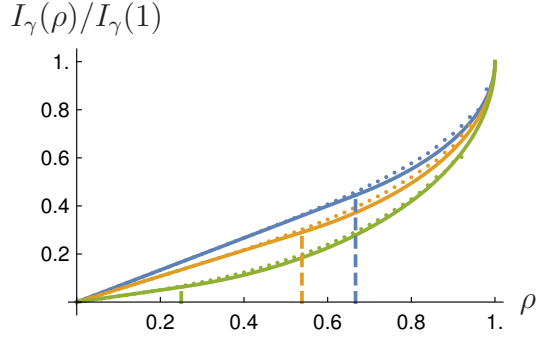}};
		\begin{scope}[x={(image.south east)},y={(image.north west)}]
		\node at (0.1,1.1) {$I_\gamma(\rho)/ I_\gamma(1)$};
		\node at (1.05,0.1) {$\rho$};
		\end{scope}
		\end{tikzpicture}
\caption{\emil{The rate function for the ARW occupation fluctuation. The whole lines show exact analytical results, with the vertical dashed lines indicating the phase transition points. Dots show  simulation results of about $5\times10^{9}$ trajectories with duration $T=20$. The values for $\gamma$ used are $0.2,0.4,0.6$ (blue, yellow, green)}} \label{fig:occ_arw_rf}
\end{figure}

For the ARW, the integral to be computed in \eqref{eq:tildeZ} is
\begin{equation}
\mathcal{I} = \frac{1}{2\pi i} \oint_{\del D} \frac{dz}{z}\, \frac{1}{u - (z + \gamma z^{-1} - (1+\gamma))} = -\frac{1}{2\pi i} \oint_{\del D} \frac{dz}{z^2 - az + \gamma},
\end{equation}
with $a = u + 1+\gamma$. The poles are
\begin{equation}
z_\pm = a/2 \pm \sqrt{(a/2)^2 - \gamma}.
\end{equation}
One verifies that $z_+(u=0) = 1$ and $\del_u z_+ > 0$, hence this pole lies outside $D$ for $u>0$; $z_-(u=0) = \gamma$, $z_-(u\to\infty)=0$ whilst $\del_u z_- < 0$, hence this pole lies inside $D$ for $u>0$ and $\gamma < 1$. Using the residue theorem, 
\begin{equation}
\mathcal{I} = \frac{1}{z_+ - z_-} = \frac{1}{\sqrt{a^2 - 4\gamma}} = \frac{1}{\sqrt{u(u+2(1+\gamma)) + (1-\gamma)^2}}\; .
\end{equation}
Therefore,
\begin{equation}
\tilde{Z}(u,s) = \frac{1}{u}\cdot \left[1 + \frac{s}{\sqrt{u(u+2(1+\gamma)) + (1-\gamma)^2}}\right]^{-1}.
\end{equation}
Only if $s < s^* = - (1-\gamma)$ does there exist a positive pole, namely
\begin{equation}
u^* = - (1+\gamma) + \sqrt{s^2 + 4\gamma}\;.
\end{equation}
We therefore conclude
\begin{equation}\label{eq:Lambda_res_ARW}
\Lambda(s) = \begin{cases}
- (1+\gamma) + \sqrt{s^2 + 4\gamma}, & s < s^*,\\
0, & s \geq s^*.
\end{cases}
\end{equation}
Note that in contrast to the version of this problem in continuous space \cite{Nyawo2018}, we are able here to obtain an explicit expression for $\Lambda(s)$.

We can see directly that $\Lambda(s)$ is not analytic in $s$, which means that the G\"artner-Ellis theorem does not apply. \emil{ Operating on the assumption (to be justified) that the rate function is convex, we nonetheless obtain the rate function via \eqref{eq:I}, as explained in \autoref{sec:rf}.} In the restricted range $s \geq s^*$, the maximizer of \eqref{eq:I} is $s^*$. For $s \leq s^*$, the maximizer is instead
\begin{equation}\label{eq:sdag}
s^\dagger = - \frac{2\rho \sqrt{\gamma}}{\sqrt{1-\rho^2}}\;,
\end{equation}
provided that $s^{\dagger} < s^*$. This condition is equivalent to
\begin{equation}
\rho > \rho_c = \frac{1-\gamma}{1+\gamma}\;.
\end{equation}
 The rate function is therefore
\begin{equation}
I_\gamma(\rho) = \begin{cases}
(1-\gamma) \rho,& \rho \leq \rho_c, \\
1+\gamma - 2\sqrt{\gamma(1-\rho^2)}, & \rho > \rho_c,
\end{cases}
\end{equation}
which is plotted in \autoref{fig:occ_arw_rf}. The salient feature is the change from linear to convex shape at $\rho_c$, which reveals a dynamical phase transition (first order since the `free energy' $\Lambda(s)$ has discontinuous derivative). \emil{The correctness of the linear branch of the rate function, and hence convexity, is verified by empirical calculation of the rate function from naive sampling of a large number of simulated trajectories.} 

We remark that in the case where $\gamma = 0$ the rate function is entirely linear, which follows immediately from the Poisson distribution of leaving the origin, after which a return is impossible. This observation already suggests that the linear part for non-zero $\gamma$ is due to trajectories that stay close to the origin for the chosen time-fraction before venturing off to infinity. Simulations corroborate this statement.

\subsection{Asymmetric random walker---effective process}

\begin{figure}
	\centering
	\begin{tikzpicture}
		\node[anchor=south west,inner sep=0] (image) at (0,0){\includegraphics[scale=0.55]{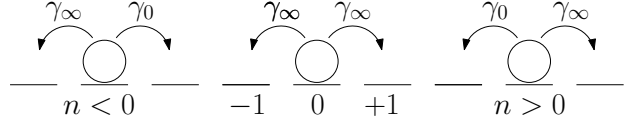}};
		\begin{scope}[x={(image.south east)},y={(image.north west)}]
		\end{scope}
	\end{tikzpicture}
		\caption{In the ARW effective process for $\rho>\rho_c$, jumps toward the origin occur with rates $\gamma_0(\rho)$ and away from the origin with rates $\gamma_\infty(\rho).$}\label{fig:rates_arw_res}
\end{figure}

The left eigenvectors $\ell(n)$ needed in \eqref{eq:Wd} to construct the transition rates of the effective process are the solutions to
\begin{equation}
\Lambda(s) \ell(n) = \ell(n+1) + \gamma \ell(n-1) - (1+\gamma)\ell(n) - s\ell(0)\delta_{n,0}\;.
\end{equation}
In the convex regime of the rate function, corresponding to $s < s^*$,
\begin{equation}
\ell(n+1) + \gamma \ell(n-1) - \sqrt{s^2 + 4\gamma}\,\ell(n) = s \delta_{n,0}\;.
\end{equation}
The solution by generating function techniques is
\begin{align}
\ell(n) 
& =  \begin{cases}
r^{-n},& n \geq 0\\
(\gamma r)^{-|n|}, & n < 0\;.
\end{cases}
\end{align}
where
\begin{equation}
r = \frac{1}{2\gamma} [ \sqrt{s^2 + 4 \gamma} - s]\;. 
\end{equation}
This leads to the effective rates
\begin{subequations}
	\begin{align}
	W^{\text{eff}}(n\to n+1) &= \begin{cases}
	\frac{1}{2}(\sqrt{s^2 + 4\gamma} + s), & n \geq 0 \\
	\frac{1}{2}(\sqrt{s^2 + 4\gamma} - s), & n < 0\;,
	\end{cases} \\	
	W^{\text{eff}}(n+1\to n) &= \gamma / W^{\text{eff}}(n\to n+1)-;.
	\end{align}
\end{subequations}
The activity parameters are preserved, but the spatial homogeneity of the original rates is lost since the observable is spatially dependent---it singles out a specific site on the lattice. To make a fluctuation $\rho \geq \rho_c$ typical, we substitute $s^\dagger$ \eqref{eq:sdag} for $s$. The resulting effective process is described by \autoref{fig:rates_arw_res} with
\begin{subequations}
	\begin{align}
	\gamma_0(\rho) = \sqrt{\gamma \frac{1+\rho}{1-\rho}}\;,\quad
	\gamma_\infty(\rho) = \sqrt{\gamma \frac{1-\rho}{1+\rho}}\;.
	\end{align}
\end{subequations}
Since $\gamma_0/\gamma_\infty > 1$, the particle is biased towards the origin, i.e.\ it lives in a linear confining potential that becomes steeper as $\rho$ is increased. Exactly at the transition point $\rho_c$, $\gamma_0 = 1$ and $\gamma_\infty = \gamma$. The original dynamics has then been modified by mirroring the dynamics for the negative half-lattice to achieve symmetry about the origin.

\begin{figure}
\begin{subfigure}{0.5\textwidth}
	\centering
	\begin{tikzpicture}
	\node[anchor=south west,inner sep=0] (image) at (0,0){
		\includegraphics[scale=1.2]{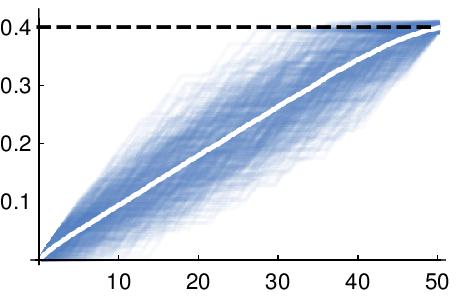}};
	\begin{scope}[x={(image.south east)},y={(image.north west)}]
	\node at (0.05,1.1) {$\mathcal{O}_t/T$};
	\node at (1.05,0.1) {$t$};
	\end{scope}
	\end{tikzpicture}
	\caption{Time at origin, $\rho > \rho_c$}
\end{subfigure}
\begin{subfigure}{0.5\textwidth}
	\centering
	\begin{tikzpicture}
	\node[anchor=south west,inner sep=0] (image) at (0,0){
		\includegraphics[scale=1.2]{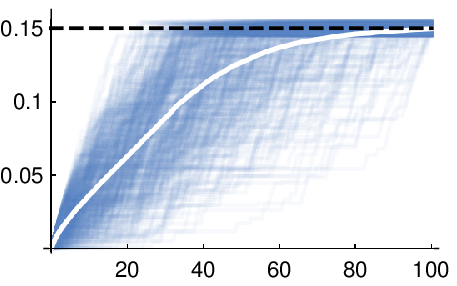}};
	\begin{scope}[x={(image.south east)},y={(image.north west)}]
	\node at (0.05,1.1) {$\mathcal{O}_t/T$};
	\node at (1.05,0.1) {$t$};
	\end{scope}
	\end{tikzpicture}
	\caption{Time at origin, $\rho < \rho_c$}
\end{subfigure}
\caption{\emil{The increase in fraction of time $\mathcal{O}_t/T$ spent at origin as a function of $t$, $0 \leq t \leq T$, from simulation of $\sim10^7$ ARW trajectories with $\gamma=0.6$. In \textbf{(a)} a subset of $500$ trajectories satisfying $\rho = 0.4\pm0.05 > \rho_c = 0.25$ at final time $T=50$  have been selected;  in \textbf{(b)} $\rho = 0.15 \pm 0.005 < \rho_c$ and $T=100$. The white line indicates the average over trajectories. For $\rho > \rho_c$ the approximate linearity of the average shows that the contribution to $\mathcal{O}(t)$ is evenly distributed throughout; the particle is localized around the origin. For $\rho < \rho_c$, there is a localized period, followed by escape.}}\label{fig:sim_arw}
\end{figure}

For $\rho < \rho_c$, since the rate function is not strictly convex, the asymptotic equivalence of the effective process and the conditioned process breaks down, as we can plainly see: \emil{attempting to substitute $s = - I'(\rho)$, we find $s = s^* = - (1-\gamma)$ for all $\rho < \rho_c$. This however would produce effective rates $\gamma_0(\rho) = 1$ and $\gamma_\infty(\rho) = \gamma$ which always gives a typical occupation $\rho_c$, and not the chosen $\rho < \rho_c$! Turning instead to direct simulation, we find, in expected agreement with \cite{Nyawo2018}, the trajectory structure shown in \autoref{fig:sim_arw}. For long simulation times $T$, the graphs show how the accumulated occupation time at site 0  increases with time for a large sample of trajectories satisfying the final occupation $\rho$ within some tolerance. Up to a finite time effect, for $\rho > \rho_c$ the contribution is evenly spread through the time interval, resulting in the average occupation increasing linearly (white line). In contrast, for $\rho < \rho_c$ the average increases linearly for about half the simulation time, at which point most of the trajectories have achieved the final occupation. Thus, these trajectories stay localized around the origin for an initial portion of time before escaping off to infinity. }
	
\subsection{Asymmetric run-and-tumble particle---rate function}

\begin{figure}[t]
	\centering	
	
	\begin{subfigure}{\textwidth}
		\centering
		\begin{tikzpicture}
		\node[anchor=south west,inner sep=0] (image) at (0,0){\includegraphics[scale=1.2]{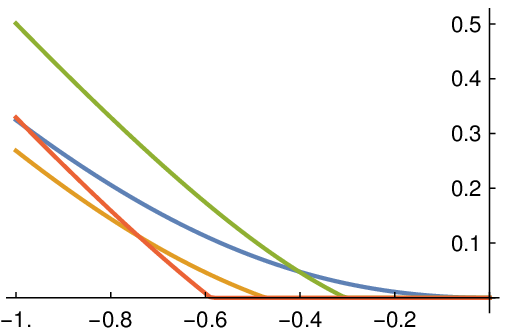}};
		\begin{scope}[x={(image.south east)},y={(image.north west)}]
		\node at (0.1,1.1) {$\Lambda(s)$};
		\node at (1.05,0.1) {$s$};
		\end{scope}
		\end{tikzpicture}
		\begin{tikzpicture}
		\node[anchor=south west,inner sep=0] (image) at (0,0){\includegraphics[scale=1.2]{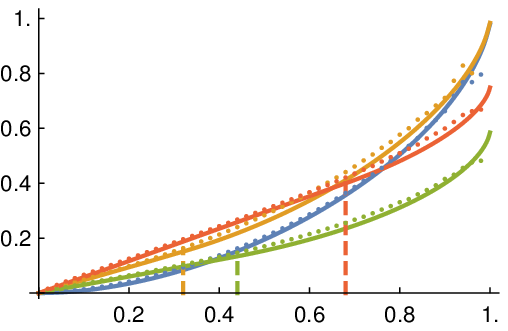}};
		\begin{scope}[x={(image.south east)},y={(image.north west)}]
		\node at (0.1,1.1) {$I(\rho)$};
		\node at (1.05,0.1) {$\rho$};
		\node at (0.2,0.65) {\includegraphics[scale=0.8]{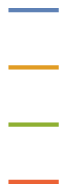}};
		\node at (0.38,0.83) {\scriptsize totally sym.};
		\node at (0.42,0.71) {\scriptsize asym.\ tumbling};
		\node at (0.41,0.59) {\scriptsize asym.\ hopping};
		\node at (0.395,0.47) {\scriptsize totally\ asym.};
		\end{scope}
		\end{tikzpicture}
		\caption{Various (a)symmetric parameter choices}
	\end{subfigure}
	\begin{subfigure}{\textwidth}
		\centering
		\begin{tikzpicture}
		\node[anchor=south west,inner sep=0] (image) at (0,0){\includegraphics[scale=1.2]{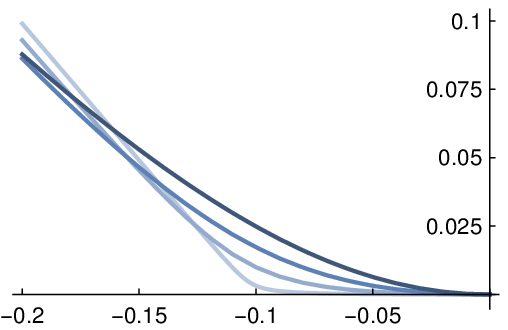}};
		\begin{scope}[x={(image.south east)},y={(image.north west)}]
		\node at (0.1,1.1) {$\Lambda(s)$};
		\node at (1.05,0.1) {$s$};
		\end{scope}
		\end{tikzpicture}
		\begin{tikzpicture}
		\node[anchor=south west,inner sep=0] (image) at (0,0){\includegraphics[scale=1.2]{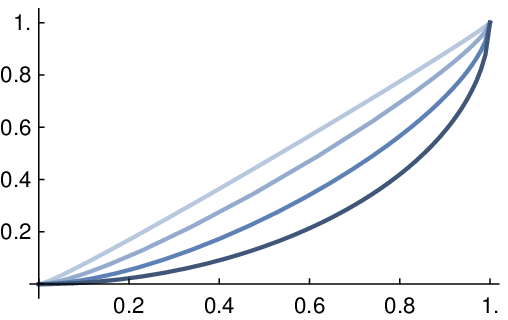}};
		\begin{scope}[x={(image.south east)},y={(image.north west)}]
		\node at (0.1,1.1) {$I(\rho)/ I(1)$};
		\node at (1.05,0.1) {$\rho$};
		\end{scope}
		\end{tikzpicture}
		\caption{Zero limiting velocity}\label{fig:res_rtp_scgf_rf_delta}	
	\end{subfigure}		
	\caption{SCGF (left panels) and rate function (right panels) for the RTP occupation problem. The SCGF is non-singular f and only if the limiting velocity is zero. \textbf{(a)} Parameters demonstrate the possible combinations of (a)symmetry ($\gamma,\omega,\delta\omega = 1,1,0;\ 1,1,0.5;\ 0.3,1,0;\ 0.3,0.65,-0.45$). Dots show simulation result of about $3\times 10^9$ trajectories with duration $T=25$. \textbf{(b)} The zero limiting velocity rate rates \eqref{eq:0vrates}, $\gamma=0.1$, and $\delta=10,.5,.1,.01$ from dark to light line colour. }\label{fig:res_rtp_scgf_rf}
\end{figure}

We now turn to the last of the four problems we consider, namely the occupation-time problem for the RTP. Recall that $\Lambda(s)$ was given by the largest positive singularity of the Laplace-transformed partition function $\tilde{Z}(u,t)$ which has the form \eqref{eq:tildeZ}. Let us write \eqref{eq:W_RTP} as
\begin{equation}
W(z) = \begin{pmatrix}
\mu(z) & \omega_+ \\
\omega_- & \nu(z)/z
\end{pmatrix},\quad \begin{split}\mu(z) &= z - 1 - \omega_+\;, \\ \nu(z) &= - z(\gamma+\omega_-) + \gamma\;. \end{split}
\end{equation}
Then
\begin{dmath}
[uI - W(z)]^{-1} =
\frac{1}{(u-\mu(z))(uz-\nu(z))-\omega_+\omega_-z}
\begin{pmatrix}
uz - \nu(z) & \omega_-z \\
\omega_+ z & z(u-\mu(z))
\end{pmatrix}.
\end{dmath}
Every element in the above has two poles given by the zeroes of the quadratic
\begin{equation}
(u-\mu(z))(uz-\nu(z))-\omega_+\omega_-z = -a_-(z - z_+)(z-z_-)\;,
\end{equation}  
where
\begin{subequations}\label{eq:a+-c}
	\begin{align}
	z_\pm &= \frac{1}{2 a_-} \left[ a_+a_- + \gamma - \omega_+\omega_- \pm \sqrt{(a_+a_-+\gamma - \omega_+\omega_-)^2 - 4\gamma a_+a_-} \right], \\
	a_+ &= u + 1 + \omega_+\;,\\
	a_- &= u + \gamma + \omega_-\;.
	\end{align}
\end{subequations}
One can verify that $z_- \in D\setminus \del D$ (open complex unit disk) and $z_+\notin D$ for the relevant parameter ranges and $u>0$ (in fact, both roots are real, $z_+ > 1$, and $1>z_->0$). Then by means of the residue theorem and after some algebraic simplifications we find
\begin{equation}\label{eq:calI}
\mathcal{I} = \frac{1}{2\pi i}\oint_{\del D} \frac{dz}{z}\, [uI - W(z)]^{-1} = \frac{1}{a_-(z_+-z_-)} \begin{pmatrix}
a_- - \frac{\gamma}{z_+} & \omega_- \\
\omega_+ & a_+ - z_-
\end{pmatrix},
\end{equation}
and
\begin{equation}
\tilde{Z}(u,s) = \frac{1}{u} \begin{pmatrix}1 & 1 \end{pmatrix}(I + s\mathcal{I})^{-1}\begin{pmatrix}
1\\1
\end{pmatrix}.
\end{equation}
Taking into account that $z_\pm > 0$ for $u>0$, one can conclude that the positive poles of $\tilde{Z}(u,s)$ are given by $\det (I + s\mathcal{I}) =0$, i.e. 
\begin{equation}\label{eq:detPole}
\det \begin{pmatrix}
a_-(z_+-z_-) + s(a_- - \frac{\gamma}{z_+}) & s\omega_- \\
s\omega_+ & a_-(z_+-z_-) + s(a_+ - z_-)
\end{pmatrix} = 0\;.
\end{equation}
This equation can be solved numerically for the largest pole $u^* = \Lambda(s)$. \autoref{fig:res_rtp_scgf_rf} shows that the resulting SCGF becomes zero for $s$ larger than some $s^*< 0$ in a non-differentiable fashion when the velocity is non-zero. 
Assuming the validity of \eqref{eq:I}, the rate function develops a linear-convex transition corresponding to the singularity of the SCGF.

\emil{We note that the special ARW-RTP velocity large deviation equivalence at the parameter values $\omega_+ = \gamma$ and $\omega_- = 1$ does not extend to the occupation observable. By putting $u = 0$ in \eqref{eq:detPole} and solving for $s$, we get $s^* = - (1+\gamma)(1-\sqrt{\gamma})$, different from the ARW. Since the SCGFs do not have the same singularity they surely do not coincide.}
 
When we choose rates according to \eqref{eq:0vrates} producing zero limiting velocity, we obtain \autoref{fig:res_rtp_scgf_rf_delta}. As $\delta$  becomes small, the SCGF approaches a singularity at $-\gamma$. Meanwhile, the rate function evolves continuously from convex to linear. As tumble events become rare, the time spent at the origin is dominated by the time the particle sits there before its first exit.

\subsection{Biased run-and-tumble particle---effective process}

\begin{figure}
	\centering
	\begin{subfigure}{\textwidth}
		\centering
		\begin{tikzpicture}
		\node[anchor=south west,inner sep=0] (image) at (0,0){\includegraphics[scale=.9]{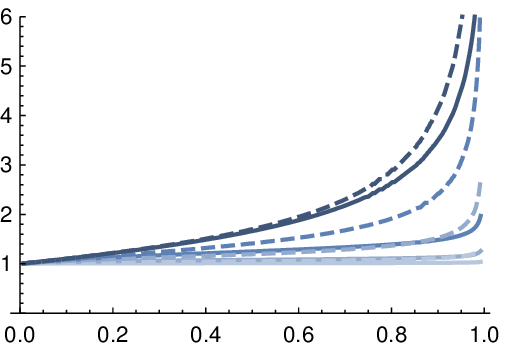}};
		\begin{scope}[x={(image.south east)},y={(image.north west)}]
		\node at (0.4,0.7) {\includegraphics[scale=.9]{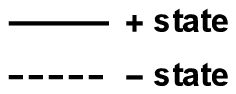}};
		\node at (-0.07,0.9) {$\tilde{\gamma}$};
		\end{scope}
		\end{tikzpicture}
		\begin{tikzpicture}
		\node[anchor=south west,inner sep=0] (image) at (0,0){\includegraphics[scale=.9]{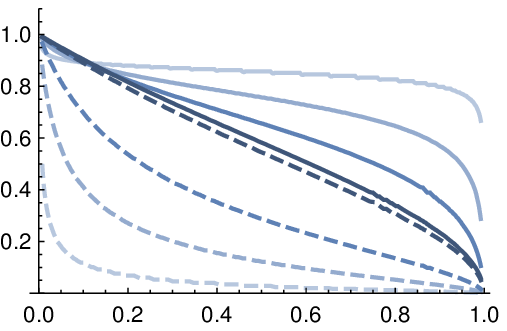}};
		\begin{scope}[x={(image.south east)},y={(image.north west)}]
		\end{scope}
		\end{tikzpicture}
		\begin{tikzpicture}
		\node[anchor=south west,inner sep=0] (image) at (0,0){\includegraphics[scale=.9]{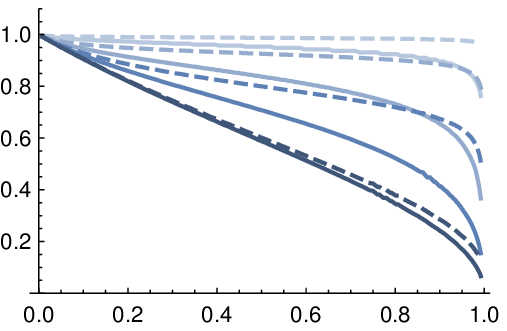}};
		\begin{scope}[x={(image.south east)},y={(image.north west)}]
		\node at (1.05,0.07) {$\rho$};
		\end{scope}
		\end{tikzpicture}
		$n<0\hspace{3.5cm} n=0 \hspace{4cm} n >0 $ 
		\caption{Effective hopping rates}
	\end{subfigure}
	\begin{subfigure}{\textwidth}
		\centering
		\begin{tikzpicture}
		\node[anchor=south west,inner sep=0] (image) at (0,0){\includegraphics[scale=.9]{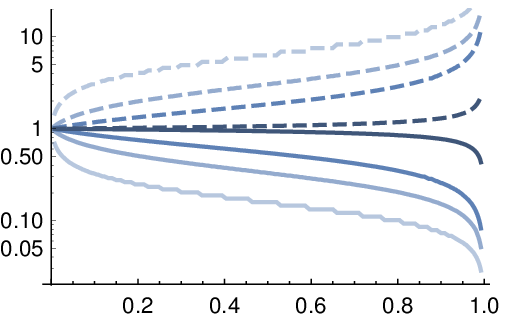}};
		\begin{scope}[x={(image.south east)},y={(image.north west)}]
		\node at (-0.03,0.9) {$\tilde{\omega}$};
		\end{scope}
		\end{tikzpicture}
		\begin{tikzpicture}
		\node[anchor=south west,inner sep=0] (image) at (0,0){\includegraphics[scale=.9]{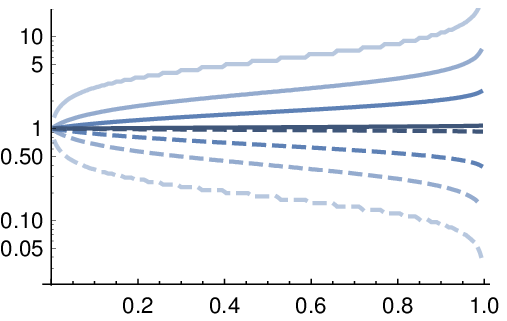}};
		\begin{scope}[x={(image.south east)},y={(image.north west)}]
		\end{scope}
		\end{tikzpicture}
		\begin{tikzpicture}
		\node[anchor=south west,inner sep=0] (image) at (0,0){\includegraphics[scale=.9]{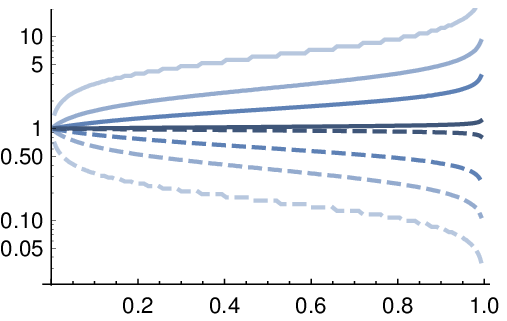}};
		\begin{scope}[x={(image.south east)},y={(image.north west)}]
		\node at (1.05,0.07) {$\rho$};
		\end{scope}
		\end{tikzpicture}\\
		$n<0\hspace{3.5cm} n=0 \hspace{4cm} n >0 $ 
		\caption{Effective tumbling rates}
	\end{subfigure}
	\caption{\emil{Effective rates corresponding to \autoref{fig:res_rtp_scgf_rf_delta} as a function of $\rho$, normalized by the original rates. Top row \textbf{(a)}: hopping rates; bottom row \textbf{(b)}: tumbling rates; whole line: $+$ state; dashed line: $-$ state. The tumbling rates are plotted on a vertical log-scale. See main text for interpretation.}}\label{fig:6rates}		
\end{figure}

To find the effective process we again seek the dominant left eigenvector $\ell_\pm(n)$ satisfying
\begin{subequations}\label{eq:ell_rtp_res}
	\begin{align}
	\Lambda(s)\ell_+(n) &= \ell_+(n+1) + \omega_+ \ell_-(n) - (1+\omega_++s \delta_{n,0}) \ell_+(n) \label{eq:ell_rtp_res_a} \\
	\Lambda(s) \ell_-(n) &= \gamma \ell_-(n-1) + \omega_- \ell_+(n) - (\gamma+\omega_-+s\delta_{n,0}) \ell_-(n) .\label{eq:ell_rtp_res_b}
	\end{align}
\end{subequations}
It was found numerically that $\Lambda(s) > 0$ for $s$ smaller than some $s^*$, and zero above it. We consider the former case, corresponding to the convex branch of the rate function. We reuse the parameters $a_\pm,z_\pm$ of \eqref{eq:a+-c}, but with $u^* = \Lambda(s)$ substituted for $u$ in their definitions. The generating function form of \eqref{eq:ell_rtp_res} is
\begin{equation}
\begin{pmatrix}
z^{-1} - a_+ & \omega_+ \\
\omega_- & \gamma z - a_-
\end{pmatrix}
\hat{\boldsymbol{\ell}}(z)
= s\boldsymbol{\ell}(0)\;.
\end{equation}
Inverting the matrix,
\begin{equation}
\hat{\boldsymbol{\ell}}(z) = \frac{s z}{a_+ \gamma (z-\tilde{z}_-)(z-\tilde{z}_+)} \begin{pmatrix}
a_- - \gamma z & \omega_+ \\
\omega_- &  a_+ - z^{-1} 
\end{pmatrix}\boldsymbol{\ell}(0)\;,
\end{equation}
with
\begin{equation}
\tilde{z}_\pm \equiv (a_-/\gamma a_+) z_{\pm} = 1/z_\mp
\end{equation}
following from $z_+z_- = \gamma a_+/a_-$. Applying the inverse transform,
\begin{gather}\label{eq:elln}
\boldsymbol{\ell}(n) =  -s  Q(n) \boldsymbol{\ell}(0)\;,\\
Q(n) \equiv -\frac{1}{2\pi i} \oint_{\del D} \frac{dz}{z^n} \frac{1}{\gamma a_+(z - \tilde{z}_+)(z-\tilde{z}_-)} \begin{pmatrix}
a_- - \gamma z & \omega_+ \\
\omega_- &  a_+ - z^{-1} 
\end{pmatrix}.
\end{gather}
Since $\Lambda(s) > 0$, we have $\tilde{z}_- \in D\setminus\del D$, while $\tilde{z}_+\notin D$. For $n=0$ we get the eigenvalue equation
\begin{gather}
[I + sQ(0)]\boldsymbol{\ell}(0) = \boldsymbol{0},\label{eq:Q0eq}
\end{gather}
with $Q(0)^\top = \mathcal{I}|_{u=\Lambda(s)}$ of \eqref{eq:calI}. We require a non-trivial solution, which implies that $I + s\mathcal{I}$, whose inverse appears in the dynamical partition function \eqref{eq:tildeZ}, is singular at $u = \Lambda(s)$. This indeed follows from the fact this value of $u$ is a pole of $\tilde{Z}$ (see \eqref{eq:detPole}). We may choose 
\begin{equation}
\boldsymbol{\ell}(0) = \begin{pmatrix}
a_-(z_+ - z_-) + s(a_+ - z_-) \\ - s\omega_-
\end{pmatrix}.
\end{equation}
For $n \geq 1$ we find instead
\begin{equation}
Q(\pm n) = \frac{(z_\mp)^{\pm n}}{a_-(z_+ - z_-)} \begin{pmatrix}
a_- - \frac{\gamma}{z_\mp} & \omega_+ \\
\omega_- & a_+ - z_\mp
\end{pmatrix}.
\end{equation}
It is then a matter of simple algebra to find the relevant ratios of eigenvector components, which subsequently produce the effective rates
\begin{subequations}
	\begin{gather}
	W^{\text{eff}}(+,n \to +,n+1) = \begin{cases}
	z_- & n > 0\\
	z_- \frac{a_-}{a_- + s} \frac{a_+ - z_+}{a_+ - z_-} & n = 0 \\
	z_+ & n < 0
	\end{cases}\\
	W^{\text{eff}}(-,n \to -,n-1) = \begin{cases}
	\gamma/z_+  & n > 0\\
	(\gamma/z_+) \frac{a_- + s}{a_-} & n = 0 \\
	\gamma/z_- & n < 0
	\end{cases}\\
	W^{\text{eff}}(+,n \to -,n) = \begin{cases}
	\omega_+\omega_- \frac{a_+}{a_-(a_+ - z_+)} & n > 0\\
	\omega_+\omega_- \frac{a_+}{(a_-+s)(a_+ - z_-)} & n = 0\\
	\omega_+\omega_- \frac{a_+}{a_-(a_+ - z_-)} & n <  0
	\end{cases}\\
	W^{\text{eff}}(-,n \to +,n) = \omega_+\omega_- / W^{\text{eff}}(+,n \to -,n)\;.
	\end{gather}
\end{subequations}

In contrast to the ARW, the activity parameters are not unchanged by the effective process (except for the tumbling transitions). Furthermore, the particle has a special set of rates at the origin. If $\bar{v} = 0$, then the above solution is fully sufficient as the maximizer $s(\rho) \leq s^* = 0$. \autoref{fig:6rates} shows the effective rates corresponding to the rates function \autoref{fig:res_rtp_scgf_rf_delta}. In the original process the particle hops faster to the right, but spends more time in the left-oriented state. For $n < 0$, the effective process increases both hopping rates, while increasing tumbling frequency into the right-moving state, and decreasing the tumbling frequency out of it. For $n > 0$, all hopping rates are decreased, while tumbles into the left-moving state becomes relatively more favoured. A more remarkable result is that at the origin, the left hopping rates are decreased dramatically, even for small $\rho$, and the tumbles into the left-moving states are simultaneously increased. Overall, away from the origin the effective process generates a bias towards it, and at the origin the likelihood of being in the left state is increased by an order of magnitude, and its hopping rate decreased by and order of magnitude, effectively trapping it in an inactive state at the origin. 

\emil{Finally, we investigate by simulation the structure of trajectories in the linear part of the rate functions. The contribution to the occupancy as a function of time is qualitatively the same as for the ARW, \autoref{fig:sim_arw} (and hence the graph is not here reproduced). }

\section{Summary and conclusion}\label{sec:conc}

In this work our aim was to establish basic large-deviation properties for the contrasting cases of a passive and an active particle, represented by an asymmetric random walker (ARW) and run-and-tumble particle (RTP) respectively. The essential difference between these two processes is that the RTP is a two-state process exhibiting a persistence of motion. We considered two of the simplest dynamical observables: the particle velocity and the fraction of time spent at the origin.

In the case of velocity fluctuations, we obtained analytical expressions for the rate functions and the effective process for both models. Surprisingly, we found that there exists a choice of non-trivial rates for the RTP under which its rate function becomes identical to that of an ARW. By non-trivial rates, we mean that they do not make RTP model itself identical to an ARW, as happens when the tumbling rates tend to infinity. Specifically, the persistence property of the RTP, manifested as the direction of successive hops being correlated random variables, continues to distinguish it from the ARW even when the rate functions coincide. Furthermore, this equivalence is not explained by that the fact that RTPs behave diffusively in the long time limit, as that is true for every set of parameters, not just this fine-tuned choice. We further observe that for the spectrum of the RTP's Markov matrix, at these special rates half of the eigenvalues are $L$-fold degenerate, while the other half exactly coincides with the spectrum of the ARW. \emil{This suggests that these two processes share a deeper connection than large deviation equivalence. Presumably, the symmetry of the RTP transition graph, with constant escape rates equal to those of the ARW, is key. } 
It would be interesting to establish whether a whole class of models exists where this same large deviation equivalence is manifest, and if so, what the unifying feature of this class might be.

A feature of the RTP, that is absent in the ARW, is the possibility of time-scale separation. This is obtained by letting the tumbling rates become much larger or smaller than the hopping rates, while keeping left and right tumbling rates at a fixed ratio. Specifically, we may chose this ratio such that the limiting velocity is zero. For fast tumbling rates, the rate function can be made approximately symmetric around zero, while the mechanism generating a positive or negative fluctuations differs, as the rates are highly asymmetric but in a complementary way. In the other limit of small tumbling rates, a flat region develops with a finite width, and as the tumbling rates tend to zero, the rate function approaches a singularity at endpoints of this region. This flat region is due to fluctuations in the times between tumbles being amplified by the relatively rapid hopping events. However, sending transition rates to zero sends the associated time-scale of the transition to infinity, an operation which does not necessarily commute with the long-time limit of the dynamical large deviation theory. Care therefore has to be taken in interpreting such singular limits \cite{Whitelam2018}.

For the occupation-time fluctuations, our main conclusion is the robustness of the dynamical phase transition observed by Nyawo and Touchette\cite{Nyawo2018} for biased Brownian motion in continuous 1D space. One advantage of studying this observable on the lattice (as opposed to the continuum) is that we were able to obtain an explicit expression for the rate function and \emil{effective process (within the limits of applicability)}. The ensuing analysis led, as expected, to the same conclusions, namely that if the particle has a hopping bias, the rate function exhibits a transition from linear to convex at a critical time-fraction $\rho_c$. Simulations confirm that in the convex regime, trajectories stay localized around the origin for their entire duration, whereas in the linear regime they escape the origin after some fraction of the total time.  

We further established that the same picture carries over to the RTP, although the complications of dealing with internal states meant settling for numerically exact results for the rate function and critical time-fractions, based on implicit formulas. For the RTP, it is however possible to achieve zero net velocity (hence recurrence) by an asymmetric choice of rates, and we indeed observed the linear-convex transition only when the net velocity was non-zero, making the original process transient. We believe then that, in support of the hypothesis of \cite{Nyawo2018}, transience and a non-symmetrizable Markov matrix are the key ingredients. We found for the RTP specifically that 
a non-zero velocity was a necessary and sufficient condition for the transition to take place.
It is worthwhile noting that for the transition to occur, breaking detailed balance is necessary but not sufficient, as the RTP trivially breaks detailed balance but may still have zero velocity. 

In addition to a deeper understanding of the large deviation equivalence of the ARW and RTP velocity fluctuations at special parameter values, it would be satisfying to obtain a rigorous argument for the dynamical phase transition in occupation arising generally for transient walks. Contrariwise, it would be curious if a counter-example of a transient walk within our 1D class was found for which the transition was absent.

\section*{Acknowledgements}
Emil Mallmin acknowledges studentship funding from EPSRC grant no.\ EP/N509644/1.

\end{document}